\documentclass[a4paper,11pt]{article}
\pdfoutput=1 
\usepackage[english]{babel}
\usepackage{jcappub} 
\usepackage{subfigure}
\usepackage[T1]{fontenc} 
\usepackage{enumerate}
\usepackage{xfrac}   
 \usepackage[dvipsnames]{xcolor}
\newcommand{\RESNOVA}{RES-NOVA }
\newcommand{\CEvNS}{CE$\nu$NS }
\title{RES-NOVA sensitivity to core-collapse and failed core-collapse supernova neutrinos}

\author[a,b,1]{L.~Pattavina\note{Corresponding authors.}}
\author[c,1]{N.~Ferreiro Iachellini}
\author[a,d]{L.~Pagnanini}
\author[c]{L.~Canonica}
\author[a,d]{E.~Celi}
\author[e,f]{M.~Clemenza}
\author[d,g]{F.~Ferroni}
\author[e,f]{E.~Fiorini}
\author[c]{A.~Garai}
\author[e,f]{L.~Gironi}
\author[c]{M.~Mancuso}
\author[a]{S.~Nisi}
\author[c]{F.~Petricca}
\author[a]{S.~Pirro}
\author[e,f]{S.~Pozzi}
\author[a,d]{A.~Puiu}
\author[b]{J.~Rothe}
\author[b]{S.~Sch{\"o}nert}
\author[c]{L.~Shtembari}
\author[b]{R.~Strauss}
\author[b]{V.~Wagner}

\affiliation[a]{INFN Laboratori Nazionali del Gran Sasso, Via G. Acitelli 22, I-67100 Assergi, Italy}
\affiliation[b]{Physik-Department and Excellence Cluster Origins, Technische Universit{\"a}t M{\"u}nchen, James-Franck-Stra{\ss}e 1, DE-85747 Garching, Germany}
\affiliation[c]{Max-Planck-Institut f{\"u}r Physik,  F{\"o}hringer Ring 6, DE-80805 M{\"u}nchen, Germany}
\affiliation[d]{Gran Sasso Science Institute,  Vaile F. Crispi 7, I-67100 L’Aquila, Italy}
\affiliation[e]{INFN Sezione Milano - Bicocca, Piazza della Scienza 3, I-20126 Milano, Italy}
\affiliation[f]{Dipartimento di Fisica, Universit\`a di Milano - Bicocca, Piazza della Scienza 3, I-20126 Milano, Italy}

\affiliation[g]{INFN Sezione di Roma-1, P.le Aldo Moro 2, I-00185 Roma, Italy}

\emailAdd{luca.pattavina@lngs.infn.it}
\emailAdd{ferreiro@mpp.mpg.de}
\emailAdd{lorenzo.pagnanini@gssi.it}

\emailAdd{canonica@mpp.mpg.de}
\emailAdd{emanuela.celi@gssi.it}
\emailAdd{massimiliano.clemenza@mib.infn.it}
\emailAdd{ferroni@gssi.it}
\emailAdd{ettore.fiorini@mib.infn.it}
\emailAdd{garai@mpp.mpg.de}
\emailAdd{luca.gironi@unimib.it}
\emailAdd{mancuso@mpp.mpg.de}
\emailAdd{stefano.nisi@lngs.infn.it}
\emailAdd{petricca@mpp.mpg.de}
\emailAdd{stefano.pirro@lngs.infn.it}
\emailAdd{stefano.pozzi@mib.infn.it}
\emailAdd{andrei.puiu@gssi.it}
\emailAdd{johannes.rothe@tum.de}
\emailAdd{stefan.schoenert@tum.de}
\emailAdd{lolian@mpp.mpg.de}
\emailAdd{raimund.strauss@tum.de}
\emailAdd{victoria.wagner@tum.de}

\abstract{\RESNOVA is a new proposed experiment for the investigation of astrophysical neutrino sources with archaeological Pb-based cryogenic detectors. \RESNOVA will exploit Coherent Elastic neutrino-Nucleus Scattering (CE$\nu$NS) as detection channel, thus it will be equally sensitive to all neutrino flavors produced by Supernovae (SNe). \RESNOVA with only a total active volume of (60~cm)$^3$ and an energy threshold of 1~keV will probe the entire Milky Way Galaxy for (failed) core-collapse SNe with $> 3~\sigma$ detection significance. The high detector modularity makes RES-NOVA ideal also for reconstructing the main parameters (e.g. average neutrino energy, star binding energy) of SNe occurring in our vicinity, without deterioration of the detector performance caused by the high neutrino interaction rate. For the first time, distances $<3$~kpc can be surveyed, similarly to the ones where all known past galactic SNe happened. We discuss the \RESNOVA potential, accounting for a realistic setup, considering the detector geometry, modularity and background level in the region of interest.
We report on the \RESNOVA background model and on the sensitivity to SN neutrinos as a function of the distance travelled by neutrinos. }

\begin{document}
\maketitle
\flushbottom

\section{Introduction}
\label{sec:intro}
Supernovae (SNe) are among the most energetic events in the Universe. They mark the end of a star's life with an intense burst of neutrinos~\cite{Baade254,Burrows:2012ew}. Why and how massive stars explode is one of the important long-standing unsolved mysteries in astrophysics. Neutrinos are known to play a crucial role in such events~\cite{Janka:2016fox}, nevertheless our understanding is still limited due to the lack of experimental observations. The knowledge we have relies mostly on hydro-dynamical simulations of the stellar matter, where also neutrino are propagated, but a direct validation of these simulations is still missing~\cite{Mirizzi:2015eza}. A timely, high resolution and high statistics detection of these neutrinos can be decisive for the understanding of the gravitational collapse and the connected neutrino emission~\cite{Vartanyan_2020}. In fact, neutrinos and gravitational waves (GWs), carry imprints of the explosion mechanism in real time, enabling a direct access to the inner stellar core~\cite{10.1093/mnras/stw1453}. A simultaneous detection of neutrinos and GWs is considered the \textit{Holy Grail} of modern multi-messenger astronomy. 


Multiple neutrino detectors are currently operating, and scrutinizing different region of the cosmos waiting for the next SN event. These experiments can be classified into three main categories: water-based Cherenkov (WBC) detectors~\cite{Suwa_2019,Abbasi:2011ss}, liquid scintillator (LS) detectors~\cite{Cadonati:2000kq,Asakura:2015bga,An_2016} and liquid Ar (LAr) time projection chambers~\cite{Migenda:2018ljh}. They all have two common features: they run detectors with active volumes ranging from few $m^3$ to several thousands $m^3$, and they are mostly sensitive only to $\overline\nu_e$/$\nu_e$.

Coherent Elastic neutrino-Nucleus Scattering (CE$\nu$NS), discovered few years ago~\cite{Akimov:2017ade}, is an ideal channel for neutrino detection. In fact, it opens a window of opportunities for the study of neutrino properties~\cite{Freedman:1973yd,Freedman:1977xn, Drukier:1983gj}, thanks to its high interaction cross-section and its equal sensitivity to all neutrino flavors. Currently, the SN neutrino community is lacking an experimental technique highly sensitive to the full SN neutrino signal. Recently, dark matter (DM) detectors, searching for nuclear recoils induced by galactic DM particles, were proposed to detect SN neutrinos via CE$\nu$NS~\cite{Lang:2016zhv,Khaitan:2018wnf,DarkSide20k:2020ymr}, given the similarities in the expected signal (i.e. low energy nuclear recoils).

All these experimental efforts are focusing on running and commissioning large-volume monolithic detectors, and in the near future, they will have to deal with some critical issues: as the scaling to larger volumes (e.g. project costs), and even more importantly their ability to reconstruct the particle energy in high rate conditions, as for example the ones produced by nearby SN events (<3~kpc).

It is difficult to forecast when and where the next SN will occur. Though, some predictions can be made through the study of the stellar formation rate and the distribution of SN remnants in a galaxy. A comprehensive review of the various methods available for estimating the expected SN rate in our galaxy, as well as a combined analysis of these, are presented in~\cite{Rozwadowska:2021lll}. The authors obtained a rate of 1.63$\pm$0.46~SN/100~y for the Milky Way Galaxy and the Local Group. However, as also pointed out by the authors, an important aspect to be taken into consideration is that the expected rate is not uniform throughout the galaxy volume. In fact, in~\cite{Schmidt:2014nya} it is shown that in the region around 1~kpc from the Sun the expected SN rate is 5-6 times greater than the galactic mean value. Furthermore, looking at the spatial distribution of all the past galactic SNe, they all occurred in a range between 1~kpc and 4~kpc~\cite{The:2006iu}. Events occurring in such proximity demand suitable detectors, able to tolerate high neutrino interaction rates. This requirement can be challenging for large-volume detectors monolithic, as the ones which are currently operated or planned in the near future. Compact and highly modular detectors are ideally suited to fulfill this requirement.

In this work, we present the background model and expected sensitivity of a newly proposed SN neutrino observatory, the \RESNOVA~\cite{Pattavina:2020cqc} project. RES-NOVA aims at optimizing the detector sensitivity per unit of target mass, rather than scaling up. This goal will be achieved exploiting CE$\nu$NS as detection channel to its full potential, by using a high segmented array of archaeological Pb-based detectors. Pb is the only element that ensures the highest cross-section, compared to conventional neutrino detection channels, and the largest nuclear stability, for achieving low-background levels. The RES-NOVA detector will be operated as a cryogenic detector to maximize the sensitivity to low energy nuclear recoils induced by neutrino interactions. Thanks to this experimental approach RES-NOVA will reach the physical limit of SN neutrino detection.

\RESNOVA is a small volume, (60~cm)$^3$, highly modular detector array of 500 units, with sensitivity to the relevant SN neutrino emission parameters as much as the largest proposed detectors. Thanks to the unique properties of archaeological Pb and the detector configuration, \RESNOVA is able to monitor the entire Milky Way Galaxy for core-collapse and failed core-collapse SNe. Furthermore, RES-NOVA will be able to precisely reconstruct the main SN parameters (e.g. total star binding energy, average neutrino energy) from SN events occurring as close as Betelgeuse~\cite{Joyce_2020} at 0.2~kpc, without being too much affected by the detector dead-time.

The structure of the paper is as follows: Section~\ref{sec:SN} outlines the main features of SNe as neutrino sources, while in Section~\ref{sec:7s} the expected signal produced by CE$\nu$NS reactions is discussed. The detector working principles, design and expected responses are described in Section~\ref{sec:RN}. In Section~\ref{sec:bkg} and Section~\ref{sec:signals} we report on the expected background and signal rate in the detector, while in Section~\ref{sec:sens} we discuss the expected detector sensitivity for core-collapse and failed core-collapse SNe occurring very close as well as at far distances. Finally, conclusions are presented in Section~\ref{sec:conc}.

\section{Supernovae as neutrino sources}
\label{sec:SN}
Massive SNe ($>8~M_\odot$) live fast and die young. The main sequence of burning elements of a star lasts few millions of years. This comes to an end once all the elements lighter than Si are consumed. At this point, a Fe core builds up, until the core reaches the critical Chandrasekhar mass. Gravitational instabilities start to occur, and gravity overcomes the outward pressure of the fusion reactions. The increasing inward pressure and density of the core, driven by the infall of the stellar envelopes, leads to neutronization and photo-dissociation processes. The first is responsible for the first prompt production of a high intensity burst of $\nu_e$. The second is indirectly connected to the production of $\overline\nu_e$. Other neutrino flavors are also produced via neutrino bremsstrahlung, pair annihilation and neutrino-neutrino interactions~\cite{Mirizzi:2015eza}.

Once the core has reached nuclear densities, the infalling matter will rebound on the core creating an outgoing pressure wave. This eventually becomes a shock wave due to the increasing pressure behind the wave. The shock wave loses energy and stalls, before making its way out, due to the inward pressure caused by the continuously infalling matter and by the dissociation of the increasing Fe envelope. At this point, the neutrinos play a crucial role in reviving the stalling shock wave by transporting heat from the inner core of the star outward. Now, the explosion is triggered and becomes unavoidable. What will remain after such a dramatic event is a high-density core, namely a neutron star. Eventually it is also possible that, while the radius of the high density core increases, its gravitational force prevents the star from exploding, by absorbing all the outer layer of the stellar envelope (failed core-collapse). In this case a black-hole is formed and, from this point on, all the stellar ejecta are not able to overcome the gravitational force of the high density core, thus neither neutrinos nor the electromagnetic components are able to escape.

\begin{figure}[t!]
\centering
\subfigure[a][]{%
\includegraphics[width=0.5\textwidth]{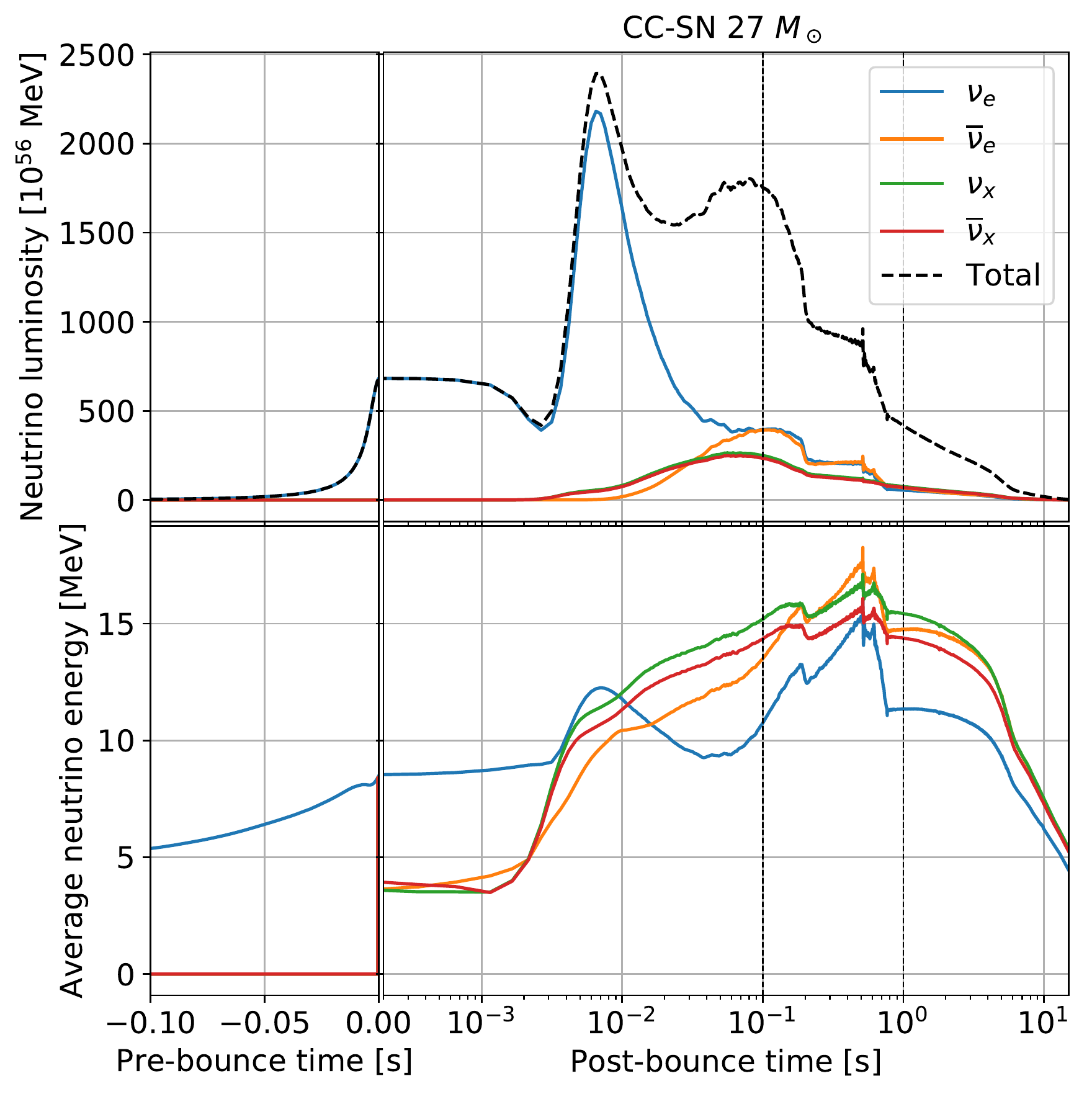}}%
\hspace{0pt}%
\subfigure[b][]{%
\includegraphics[width=0.5\textwidth]{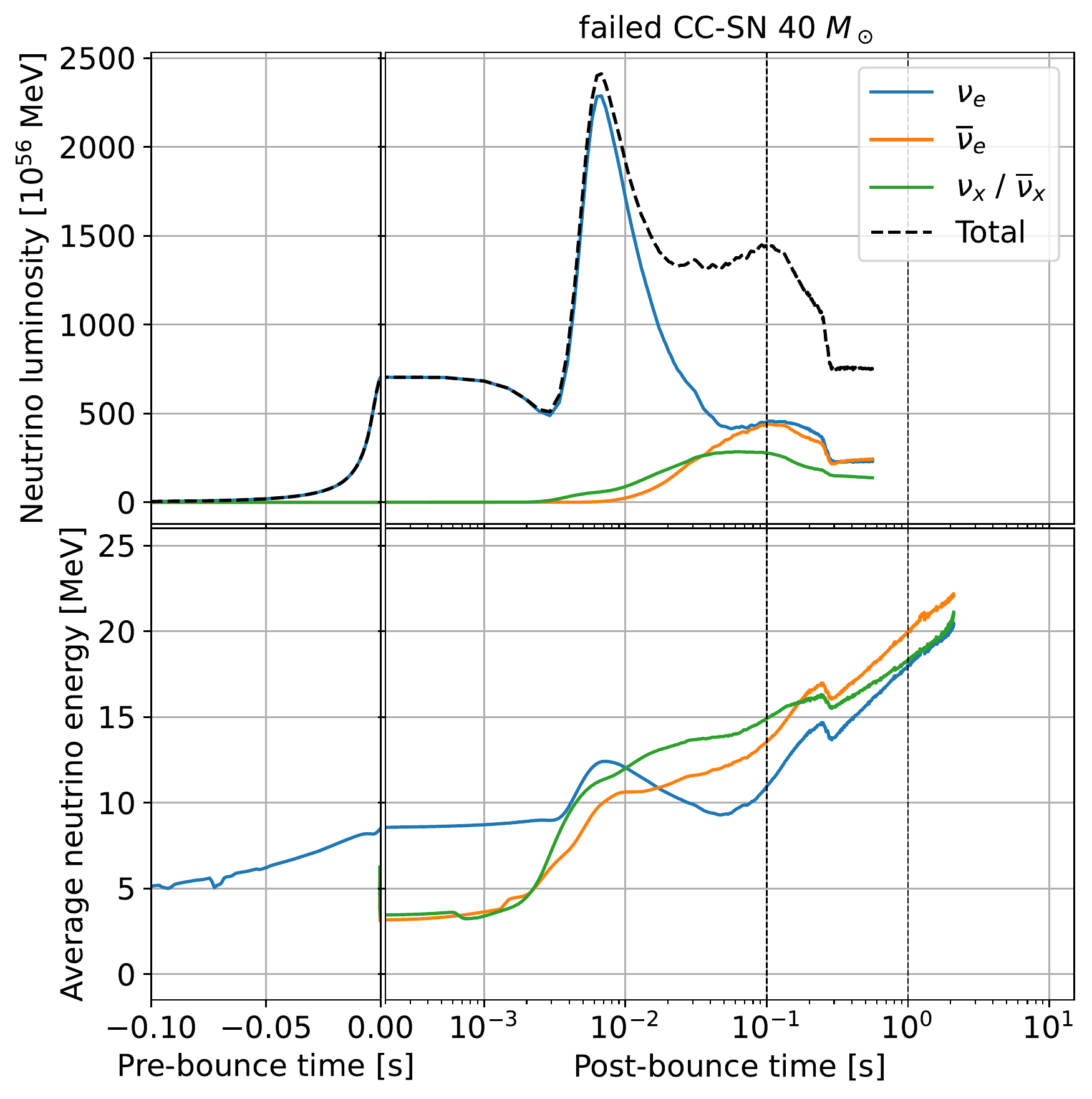}}
\caption{Time evolution of the neutrino luminosity (\textit{top row}) and average neutrino energy (\textit{bottom row}) for a core-collapse SN of $27\ M_\odot$ and a failed core-collapse SN of $40\ M_\odot$, both occurring at 10~kpc. See text for details on the models.}
\label{fig:SN_lum} 
\end{figure} 

During a SN event of the types considered in this work, three main phases can be identified: \textit{neutronization} ([0.001,0.1]~s), the shock wave is formed and it moves outwards releasing a burst of $\nu_e$; \textit{accretion} ([0.1,1]~s), neutrinos transfer energy to the outer stellar envelope revitalizing the shock wave (for a massive failed core-collapse SN this can be a longer process [0.1,2]~s, \textit{long accretion}), and \textit{cooling}, when the stellar mantle is ripped off and only a high density remnant is left (e.g. neutron star or black-hole). A time-resolved and high-statistics detection of neutrinos from these different phases will enable a deeper understanding of the mechanisms which are involved in this high energy event and at the same time will be a test bench of core-collapse physics models.
In Fig.~\ref{fig:SN_lum}), two benchmark models are shown: a core-collapse SN, $27\ M_\odot$, and a failed core-collapse, fast forming black-hole with a progenitor mass of $40\ M_\odot$. They are the same adopted in~\cite{Pattavina:2020cqc} and named \textit{LS 220} and \textit{failed-SN slow}. Given the very different nature of these SNe and the different properties of neutrinos, in the following we will adopt these as \textit{reference models}.
Looking at Fig.~\ref{fig:SN_lum}, we observe that each time window is characterized by different luminosity for each neutrino flavor, but also by different average neutrino energies. The failed-CC event can be easily identified by the sudden halt of the neutrino emission at 2~s and the constant increase of the average neutrino energy throughout its long accretion phase. They are both caused by strong gravitational force exerted by the high density core on the neutrinos. These energies when compared to the ones of solar neutrinos are almost one order of magnitude higher~\cite{Kumaran:2021lvv}. Because of these high energy processes involved, SNe are really unique high energy neutrino sources, that produce also high intensity fluxes.

An important point to be underlined is that few hours prior to the collapse, neutrinos are copiously released. This type of emission is commonly defined as \textit{pre-SN neutrinos}, and it is mostly due to the final fuel burning stage, namely Si burning~\cite{Odrzywolek:2003vn,Patton:2017neq} of the star. Such emission can be adopted as alert for the forthcoming explosion, however, the extremely small neutrino fluxes and the low energies require the operation of very large volume detectors~\cite{Simpson:2019xwo, Raj:2019wpy}, with extremely low background rates. The very last stages of the early neutrino emission are shown in Fig.~\ref{fig:SN_lum} on the negative time-axis.

\section{Coherent elastic neutrino-nucleus scattering as detection channel}
\label{sec:7s}
CE$\nu$NS was postulated in 1974~\cite{Freedman:1973yd}, but detected for the first time only in 2017~\cite{Akimov:2017ade}. The difficulties in observing this process concerned the required low detector energy threshold and the limited technology to achieve it. Thanks to the recent technological advances, CE$\nu$NS came within the reach~\cite{Schumann:2019eaa}.
The key features of this process are the high cross-section, and its neutral-current nature (i.e. a $Z^0$ is exchanged between the neutrino and the target nucleus), thus equally sensitive to all neutrino flavors. This can be used as a new portal for physics beyond the Standard Model (e.g. non-standard neutrino interactions, sterile neutrino, neutrino magnetic moment)~\cite{7s_beyondSM}, but also for the study of neutrino sources (e.g. SNe, Sun)~\cite{Drukier:1983gj}.
The total CE$\nu$NS cross-section as a function of the energy of the recoiling nucleus can be computed from Standard Model basic principles~\cite{Freedman:1973yd}: 
\begin{equation}
\label{eq:xsec}
\frac{d\sigma}{d E_R} = \frac{G^2_F m_N}{8 \pi (\hslash c )^4} \left[(4\sin^2 \theta_W -1 ) Z + N\right]^2 \left(2- \frac{E_R m_N}{E^2}\right) \cdot |F(q)|^2\ ,
\end{equation}
where $G_F$ is the Fermi coupling constant, $\theta_W$ the Weinberg angle, $Z$ and $N$ the atomic and neutron numbers of the target nucleus, while $m_N$ its mass, $E$ the  energy of the incoming neutrino and $E_R$ the recoil energy of the target. The last term of the equation, $F(q)$, is the elastic nuclear form factor at momentum transfer $q=\sqrt{2E_R m_N}$. It represents the distribution of the weak charge within the nucleus and for small momentum transfers its value is close to unity. The parameterization of $F(q)$ follows the model of Helm~\cite{Helm:1956zz}; for an exact evaluation of $F(q)$ see~\cite{PhysRevC.85.032501}.

There is a strong dependence between the recoil energy and the energy of the incoming neutrino, as shown by the average nuclear recoil energy~\cite{Drukier:1983gj}:

\begin{equation}
\label{eq:eravg}
\langle E_R \rangle = \frac{2E^2}{3m_N},
\end{equation}
ensuring a strong enhancement of the neutrino signal.

\begin{figure}
\centering
\includegraphics[width=0.7\textwidth]{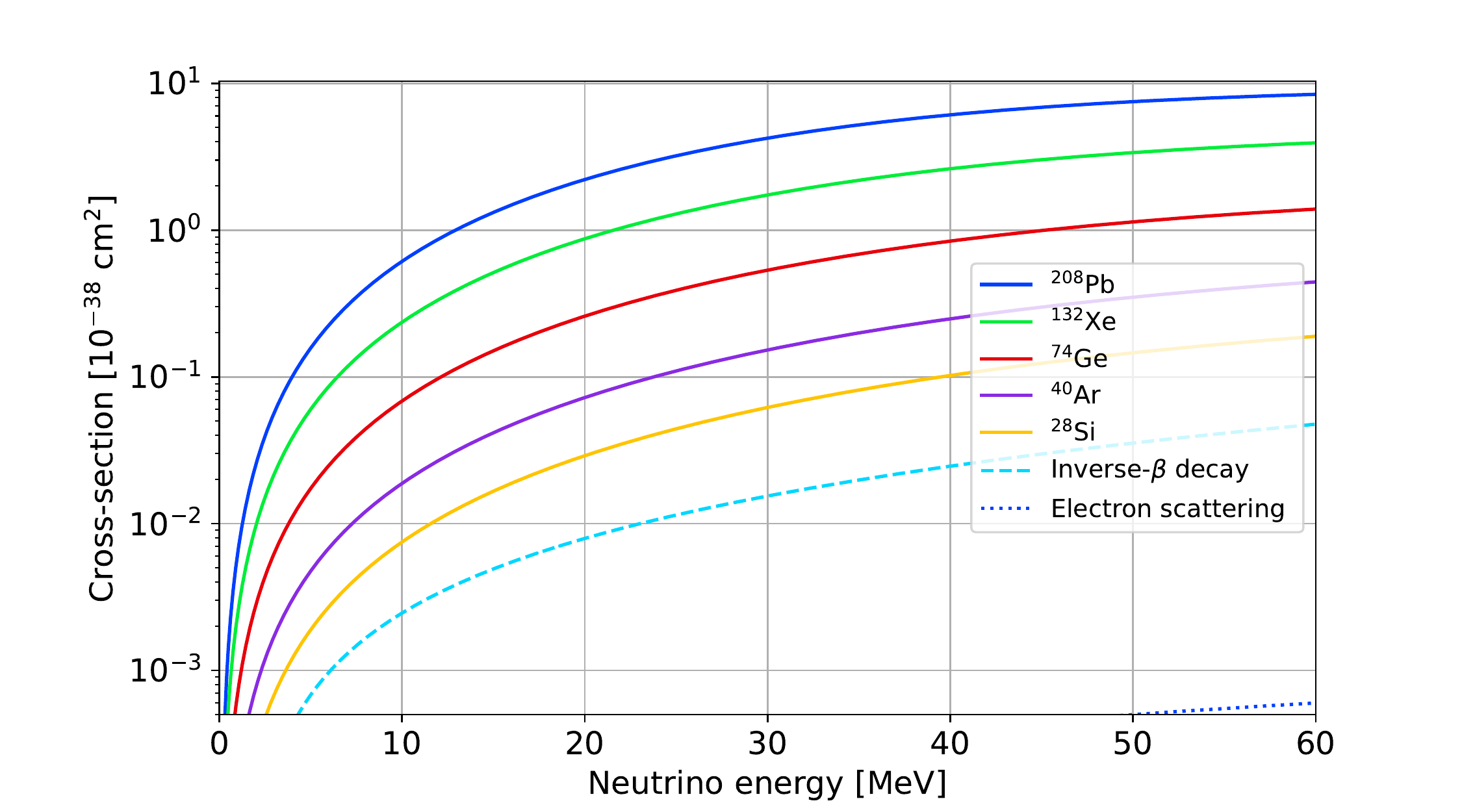}
\caption{Coherent elastic neutrino-nucleus scattering (CE$\nu$NS) cross sections as a function of the neutrino energy. The inverse-$\beta$ decay and elastic scattering on electrons cross-sections are shown as dashed and dotted lines respectively~\cite{Pattavina:2020cqc}.}\label{fig:xsec}
\end{figure}

This process has neutrino interaction cross-sections which can be $\sim 10^3$ ($\sim 10^4$) times higher than other conventional neutrino detection channels as inverse beta decay (electron scattering)~\cite{Pattavina:2020cqc}, depending on the target material. In fact, Eq.~\ref{eq:xsec} shows that having a target nucleus with a high $N$ increases $\sigma$, and if the interaction is coherent, we have a further enhancement: $\propto N^2$. In this respect, Pb can be considered as one of the best target material because it simultaneously offers the highest cross-section, for high neutrino interaction rates, and the highest nuclear stability, for ultra-low background level. In addition, there is a strong dependence of the cross-section to the energy of the incoming neutrino ($E^2$). In Fig.~\ref{fig:xsec} the neutrino interaction cross-section as a function of the neutrino energy is shown.

All these features make CE$\nu$NS an ideal channel for the detection of neutrinos of all flavors produced by high intensity sources, as nuclear reactor, the Sun or SNe. Among them, SNe are the most intense and energetic with fluxes as high as 10$^{13}~\nu$/cm$^2$/s (from a SN at 10~kpc) and energies of $\mathcal{O}(10~MeV)$~\footnote{Reactor and Solar neutrinos have fluxes of $\sim$10$^{10}~\nu$/cm$^2$/s and average energies of $\mathcal{O}(1~MeV)$~\cite{Vitagliano:2019yzm}.}. For these reasons SNe can be considered as unique neutrino sources especially when studied via CE$\nu$NS.

\section{The \RESNOVA detector}
\label{sec:RN}
\RESNOVA is a newly proposed neutrino observatory that exploits \CEvNS as detection channel and uses an array of archaeological Pb-based cryogenic detectors~\cite{Pattavina:2020cqc}. Pb is an ideal target for the detection of neutrinos from astrophysical sources via \CEvNS. In fact, it is the only element of the periodic table that ensures simultaneously the highest cross section, as this scales as the square of the neutron number of the target nucleus, and the highest nuclear stability, for achieving low-background levels. Furthermore, archaeological Pb promises unprecedented isotopic purity, leading to low background levels in the region of interest (ROI)~\cite{Alduino:2017qet, Pattavina:2019pxw}.

\RESNOVA is planned to be installed in the deep underground laboratory of Gran Sasso, where the detector can benefit from the overburden for suppressing muons and muon-induced neutron fluxes~\cite{Bellini:2012te}.
RES-NOVA research program is aiming at deploying a series of detector with increasing volumes: RES-NOVA$^1$ has a total volume of (60~cm)$^3$, compact enough to fit inside commercially available cryogenic facilities. The following upgrade is RES-NOVA$^2$ which has a volume of (140~cm)$^3$, possibly fitting inside large cryogenic infrastructures like the CUORE ones~\cite{ALDUINO20199} and ultimately RES-NOVA$^3$ which is made of 15 RES-NOVA$^2$ detectors installed in various underground facilities world-wide. In the following we will only focus on RES-NOVA$^1$, which is the first phase of the experiment and test bench for its future extensions.

\begin{figure}
\centering
\includegraphics[width=0.5\textwidth]{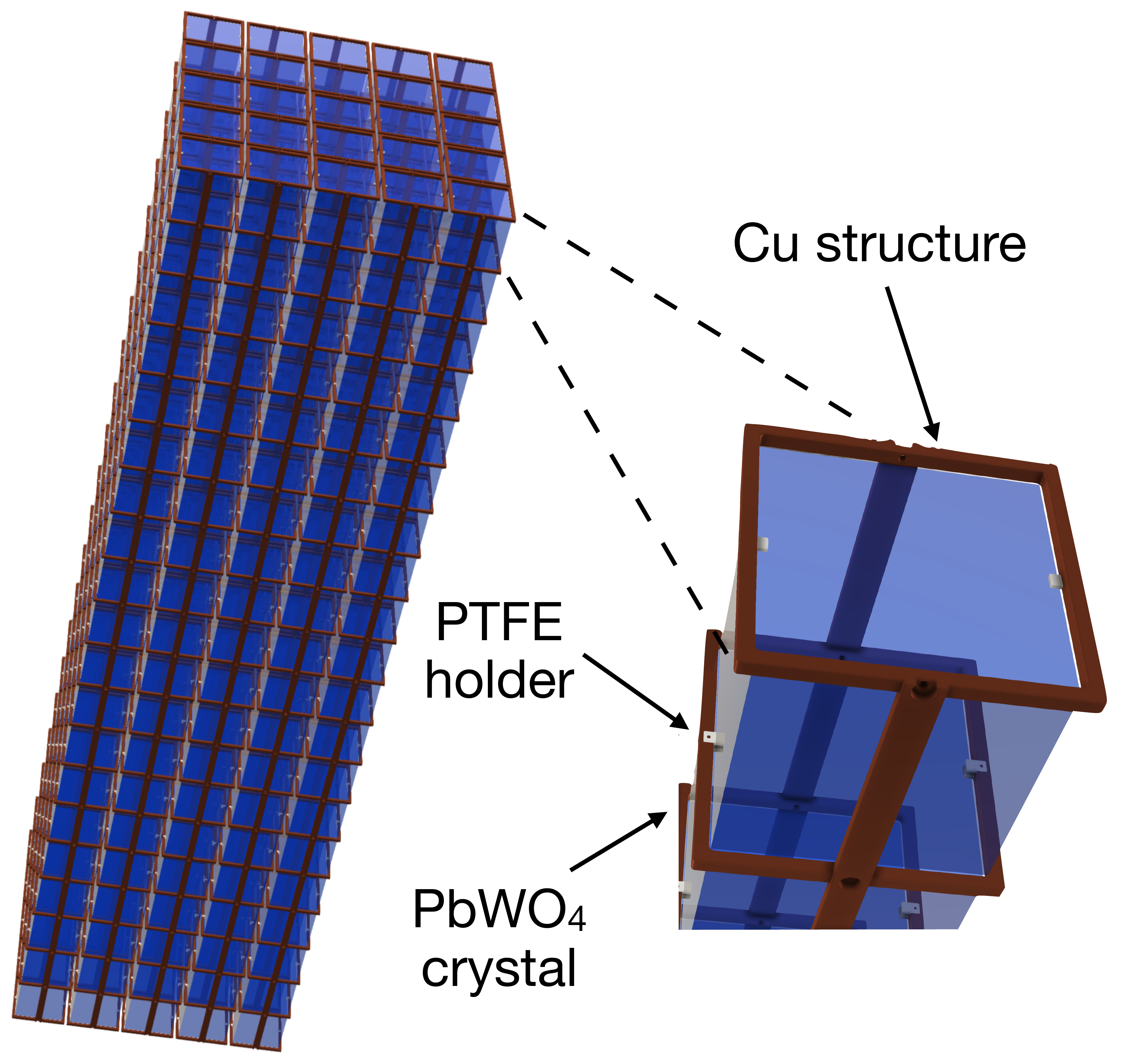}
\caption{Illustration of the detector design. The crystal absorber are arranged in a tower-like structure. On each of the 20 layers, 25 crystal are placed. The detector structure is made of Cu, while the clamps which hold the detector in place and act as thermal link are made of PTFE.}
\label{fig:detector} 
\end{figure}

The detector is composed of an array of 500 large mass Pb-based single crystals equipped with highly sensitive Transition Edge Sensor (TES) thermometers for reading out the temperature rises induced by particle interactions. This type of sensor simultaneously achieved low nuclear recoil energy thresholds ($\ll1$~keV) and fast time response $\mathcal{O}(100~\mu s)$~\cite{Schumann:2019eaa}. Each Pb-based crystal has a total volume of (7.5~cm)$^3$. They are arranged in a tower-like structure of 20 levels, each one containing 25 crystals, see Fig.~\ref{fig:detector}. This tightly packed detector configuration allows to achieve high signal-to-background ratios thanks to a coincidence data selection of the events occurring in pre-defined time windows (e.g. neutronization, accretion or cooling). The detector holding systems will be made of Cu and PTFE, with a design similar to the ones successfully adopted by the CUORE~\cite{Alduino:2017ehq} and CUPID-0~\cite{Azzolini:2019tta} experiments. The total \RESNOVA active volume is (60~cm)$^3$, a small size if compared with currently running neutrino observatories which have volumes three orders of magnitude larger~\cite{Kharusi:2020ovw}.

The crystals are operated as cryogenic calorimeters, a detector technology which demonstrated high energy resolution over a wide energy range~\cite{Pirro:2017ecr}, and most importantly limited uncertainties in event energy reconstructions due to energy quenching~\cite{Alessandrello:1999yz}. These aspects are particularly relevant for \CEvNS investigations, where the neutrino signal is expected to lie at the detector energy threshold. RES-NOVA aims at targeting energy thresholds of 1~keV, a value which is some order of magnitude more relaxed than the one currently achieved by detectors employed for direct DM investigations~\cite{Abdelhameed:2019hmk, Armengaud:2019kfj, Agnese:2017jvy, Alessandria:2012ha}. However, achieving much lower energy thresholds will not significantly enhance RES-NOVA sensitivity, as shown in~\cite{Pattavina:2020cqc}.

\RESNOVA is considering three different Pb-based crystal candidates for its investigations: pure-Pb, PbMoO$_4$ and PbWO$_4$. The best demonstrated performance were achieved with a PbWO$_4$ crystal~\cite{Beeman:2012wz}. This crystal features extremely low concentration of radioactive impurities (i.e. $^{238}$U, $^{232}$Th and $^{210}$Pb) when archaeological Pb is used, as demonstrated in~\cite{Belli:2016yof,Belli:2020qqc}. In addition, PbWO$_4$ is a widely developed crystal for high energy physics applications as major component for electromagnetic calorimeters (e.g. CMS~\cite{CMS:1997ysd} and PANDA~\cite{Erni:2008uqa}), thanks to its scintillation properties both at room and at low temperatures. Large scale production of this compound were shown to be successful~\cite{Lecoq:2005wa}. PbWO$_4$ has also demonstrated to have a light yield at low temperature which is 10$^3$ times higher than at room temperature~\cite{Lecoq:1994yr}, thus enabling a particle identification and background rejection by means of to the different light yields of interacting particles~\cite{Beeman:2012wz}. Detailed studies on the effective light yield of large volume PbWO$_4$ crystals at low temperatures are needed to properly evaluate the particle discrimination efficiency. 

\section{Detector background model}
\label{sec:bkg}
In order to deliver a robust estimate of the experimental sensitivity to SN neutrinos, the development of a detailed background model is mandatory. For this reason, starting from the current knowledge on the concentration of radioactive impurities in cryogenic low-background experiments, we developed a Monte Carlo tool for simulating the energy spectra produced by the distributions of radioactive contamination in different detector components. We can estimate the expected background level in the ROI, which lies between the detector energy threshold and 30~keV~\cite{Pattavina:2020cqc}, using as input to the Monte Carlo: the detector geometry and the concentration of background sources. 

The detector geometry described in the Monte Carlo is the one shown in Fig.~\ref{fig:detector}. We have considered all detector components next to the detector sensitive volume, that are expected to give the largest contribution to the background budget. The detector Cu holding system, the PTFE crystal supports, a vessel of 1~cm thickness and a neutron moderator of 20~cm of polyethylene define our detector geometry.

The material employed for the detector realization are: Cu, PTFE, polyethylene and PbWO$_4$. The distribution of radioactive contaminations inside the different components must be taken into account, and for this reason we simulated both bulk and surface contaminations. The latter becomes critical while dealing with cryogenic low-background experiments, given that the detector absorber is sensitive throughout its entire volume, including its surfaces~\cite{Alessandria:2012zp}.

Elements of the radioactive decay chains (i.e. $^{238}$U, $^{210}$Pb and $^{232}$Th) and environmental radioactivity (i.e. neutrons) are accountable for the largest background contributions~\cite{Abdelhameed:2019oxl,Armengaud:2013vci,Azzolini:2019nmi, Alduino:2017qet}.

\begin{table}[]
\centering
\begin{tabular}{|llll|}
\hline
\textit{Component} & \begin{tabular}[c]{@{}l@{}}\textit{Source}\\ \textit{Isotope} \end{tabular}        & \begin{tabular}[c]{@{}l@{}}\textit{Activity}\\ {[}Bq/kg{]} ({[}Bq/cm$^2${]})\end{tabular} &\\ \hline
PbWO$_4$ crystals  & $^{232}$Th             & $<2.3\times 10^{-4}$ & \cite{Belli:2020qqc} \\
                   & $^{238}$U                & $<7.0\times 10^{-5}$ & \cite{Belli:2020qqc}\\
                   & $^{210}$Pb               & $<7.1\times 10^{-4}$ & \cite{Pattavina:2019pxw}\\ 
                   \hline
Cu structure       & $^{232}$Th               & $<2.1\times 10^{-6}$ & \cite{Alduino:2017qet}\\
                   & $^{238}$U                & $<1.2\times 10^{-5}$ & \cite{Alduino:2017qet}\\
                   & $^{210}$Pb               & $<2.2\times 10^{-5}$ & \cite{Alduino:2017qet}\\
Cu surface                   & $^{232}$Th - 10 $\mu$m   & $(5.0\pm1.7)\times 10^{-9}$ & \cite{Alduino:2017qet}\\
                   & $^{238}$U - 10 $\mu$m    & $(1.4\pm0.2)\times 10^{-8}$ & \cite{Alduino:2017qet}\\
                   & $^{210}$Pb - 10 $\mu$m   & $<1.9\times 10^{-8}$ & \cite{Alduino:2017qet}\\
                   & $^{210}$Pb - 0.1 $\mu$m  & $(4.3\pm0.5)\times 10^{-8}$ & \cite{Alduino:2017qet}\\
                   & $^{210}$Pb - 0.01 $\mu$m & $(2.9\pm0.4)\times 10^{-8}$ & \cite{Alduino:2017qet}\\ \hline
PTFE holders       & $^{232}$Th               & $<6.1\times 10^{-6}$ & \cite{Alduino:2017qet} \\
                   & $^{238}$U                & $<2.2\times 10^{-5}$ & \cite{Alduino:2017qet} \\
                   & $^{210}$Pb               & $<2.2\times 10^{-5}$ & \cite{Alduino:2017qet}\\
                   \hline \hline
Environment       & neutrons              & 3.7$\times 10^{-6}$cm$^{-2}s^{-1}$ & \cite{Wulandari:2003cr} \\
                                      \hline
\end{tabular}
\caption{List of radioactive isotopes considered for the \RESNOVA background model. Values and upper limits at 90\% C.L. for the concentrations are reported for each progenitor of a decay chain. The decay chains are assumed to be in secular equilibrium. 
We considered the following specifications for the RES-NOVA detector, the total PbWO$_4$ mass: 1.8~ton; total Cu mass and surfaces: 52~kg and 1$\times$10$^6$~cm$^2$,respectively; while the PTFE holders have a total mass of 0.1~kg and have an overall surface of 2$\times$10$^3$~cm$^2$.\label{tab:material}}
\end{table}

In Table~\ref{tab:material}, there are listed the simulated background sources, their positions inside the experimental set-up and their concentrations. The numbers reported in the table are obtained from material assays through low-background detector screenings~\cite{Pattavina:2019pxw, Belli:2020qqc} and also as output of the background models of the CUORE-0~\cite{Alduino:2016vtd}, CUORE~\cite{Alduino:2017qet} and CUPID-0~\cite{Azzolini:2019nmi} experiments.
For the sake of a conservative background estimation, the limits on the concentration of radionuclides in the previously mentioned materials are taken as values for the evaluation of their contribution in the ROI.
We have classified the sources into three main categories:
\begin{itemize}
    \item Bulk contamination: the entire decay chain, starting from the progenitors, are randomly generated throughout the entire volume of the components under investigation.
    \item Surface contamination: an exponential density profile 
    of the radionuclide concentration is simulated. This profile is meant to describe a possible diffusion process of the contaminant on the outer surface of the material. The mean depth for shallow contamination is assumed to be 0.01~$\mu$m, while for the medium and deep ones we used 0.1~$\mu$m and 10~$\mu$m.
    \item External sources: background contributions induced by particles coming from outside the experimental set-up (e.g. environmental neutrons) reaching the detector.
\end{itemize}

Surface contamination are generated on all Cu components, namely the Cu structure and the inner vessel. These make up the largest surface directly facing the crystals, with a total of $4.9\times10^4$~cm$^2$ and $4.6\times10^4$~cm$^2$ respectively. The PTFE holders have a total surface of $2.0\times10^3$~cm$^2$ and radiopurity level comparable to the one of Cu, thus contributing only to a small fraction to the overall surface background budget. For this reason, we have only considered their bulk contribution.

Neutrinos of astrophysical origins (i.e. Solar neutrinos) are not taken into account for their energies are too low to contribute in the ROI. From Eq.~\ref{eq:eravg}, we can quickly estimate that the average recoil energy is roughly $\sfrac{1}{100}$ of the one expected to be produced by SN neutrinos (few keVs), well below the target threshold of the experiment (1~keV). An accurate calculation of the expected rate from Solar neutrinos for recoil detection thresholds of 1~eV, 100~eV and 1~keV leads respectively to: 10$^{-4}$~events/ton/s, 10$^{-5}$~events/ton/s and 10$^{-7}$~events/ton/s. Such rates are some orders of magnitude lower than the ones produced by radioactive decay chains.

In our background model, we have not included possible contributions from external high-energy gammas, as these are expected to be effectively suppressed by means of a Cu/Pb shielding outside the experimental set-up and will give minimal contributions at the relevant energy scales~\cite{Strauss:2014aqw,Armengaud:2013vci}. Additionally, possible cosmogenically activated nuclides in PbWO$_4$ are not taken into account due to the lack of suitable literature data. In the near future, we are planning to address these issues.

The simulations are run with a Monte Carlo code (\textit{Arby}) based on the \texttt{GEANT4} toolkit~\cite{Agostinelli:2002hh}. The output of the simulations provide a list of events releasing energy in the crystals and the relative time at which the interaction occurred. In order to take into account the detector response, the simulation output is processed with a custom made code, which smears the energy distribution of the events according to the detector energy resolution. In addition, the detector time response is also considered such that: events depositing energy in the same crystal in a specific time windows (detector time resolution) are detected as a single event with an energy that is the sum of the individual energy depositions. The output of the Monte Carlo simulations provide also a variable describing the multiplicity of an event. This is defined as the number of triggered detectors in a fixed time window (e.g. SN emission phases), as an example events like neutron or multi-Compton scattering are expected to have higher multiplicity than alpha-decays. This variable allows us to properly evaluate the background level for a given SN signal multiplicity over the same time interval.

\begin{figure}%
\centering
\subfigure[a][Neutronization ${[0.001,0.1]}$~s - $M=1$]{%
\label{fig:CC-a}%
\includegraphics[width=0.48\textwidth]{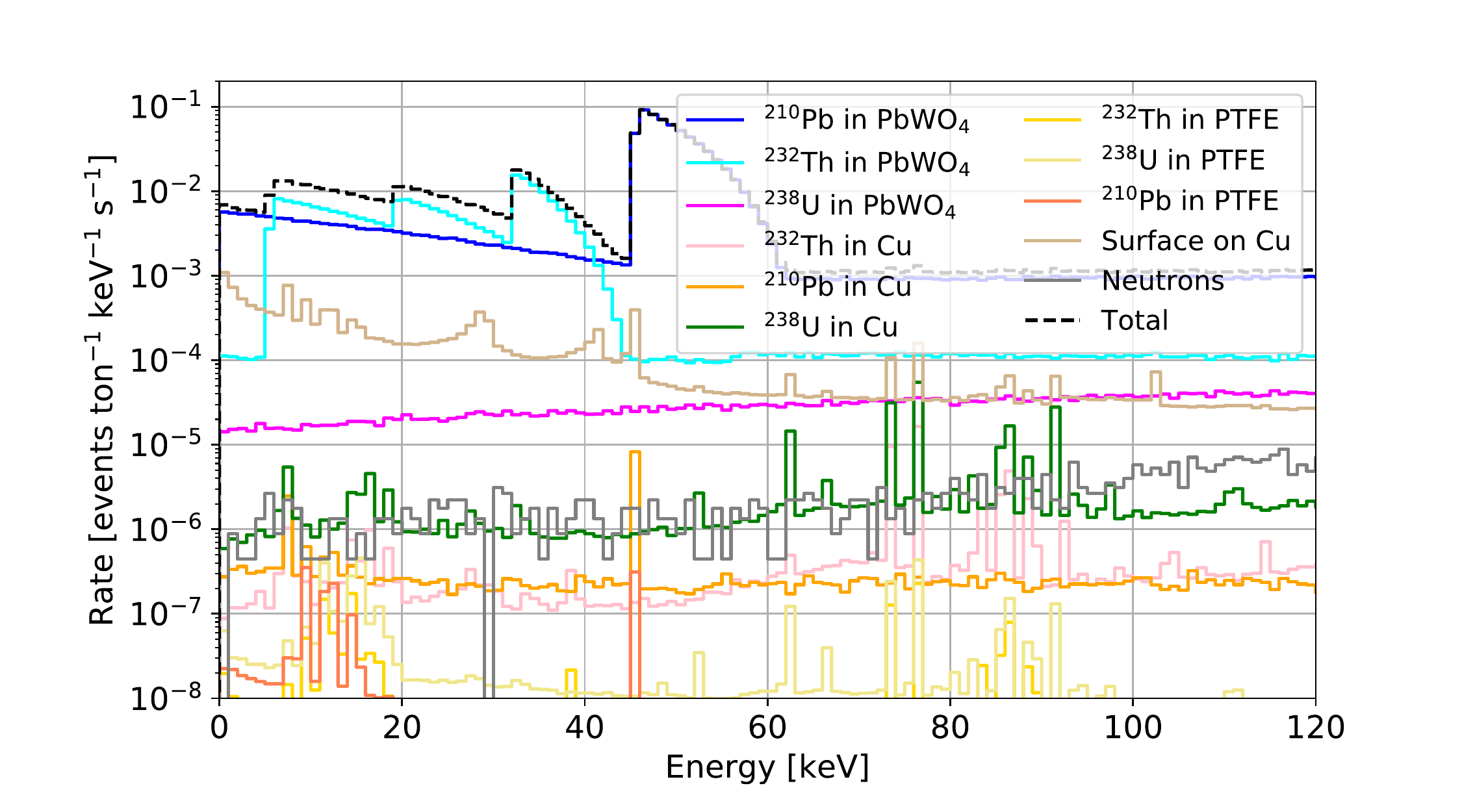}}%
\hspace{0pt}%
\subfigure[b][Neutronization ${[0.001, 0.1]}$~s - $M=4$]{%
\label{fig:CC-b}%
\includegraphics[width=0.48\textwidth]{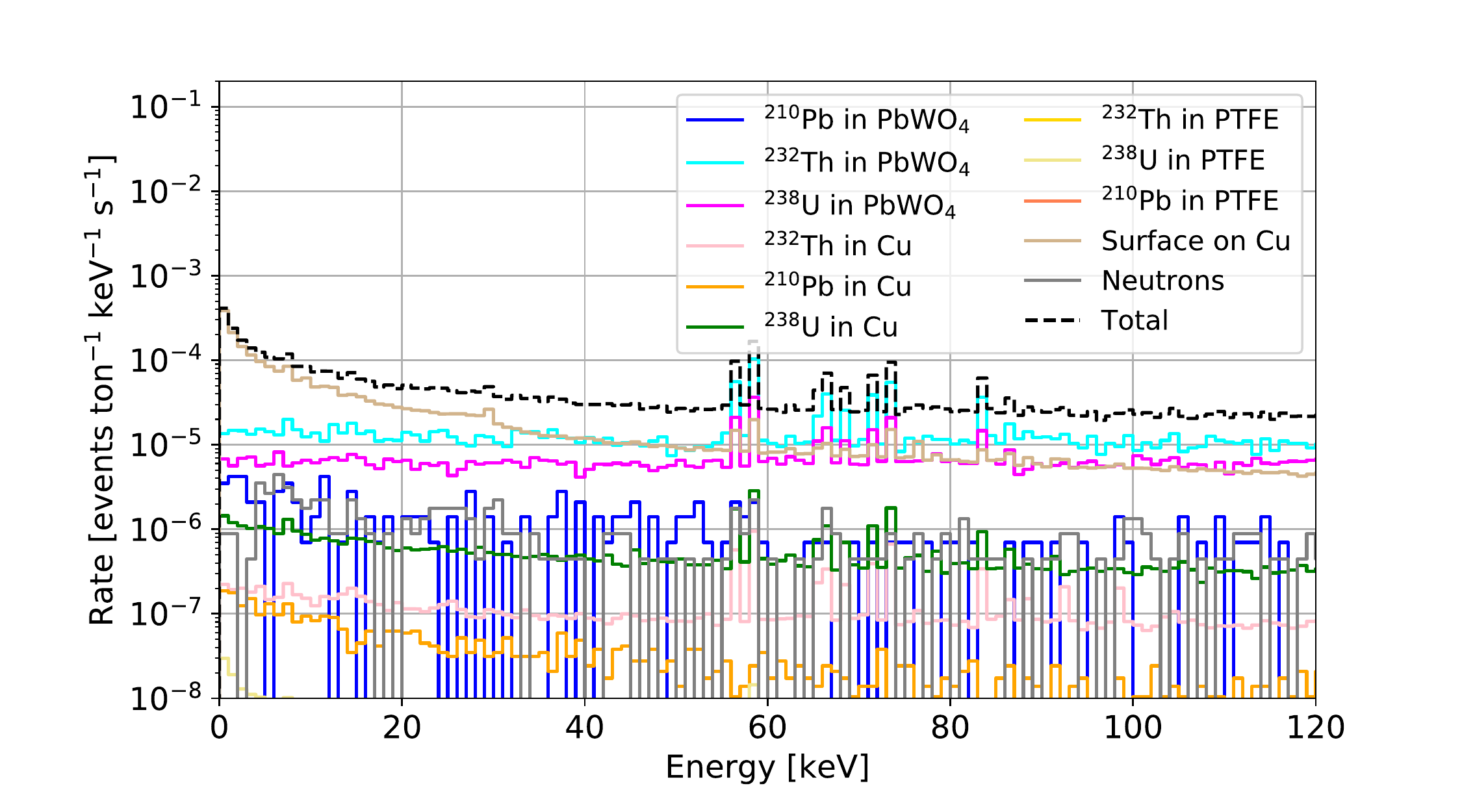}}\\
\subfigure[c][Accretion ${[0.1,1]}$~s - $M=1$]{%
\label{fig:CC-c}%
\includegraphics[width=0.48\textwidth]{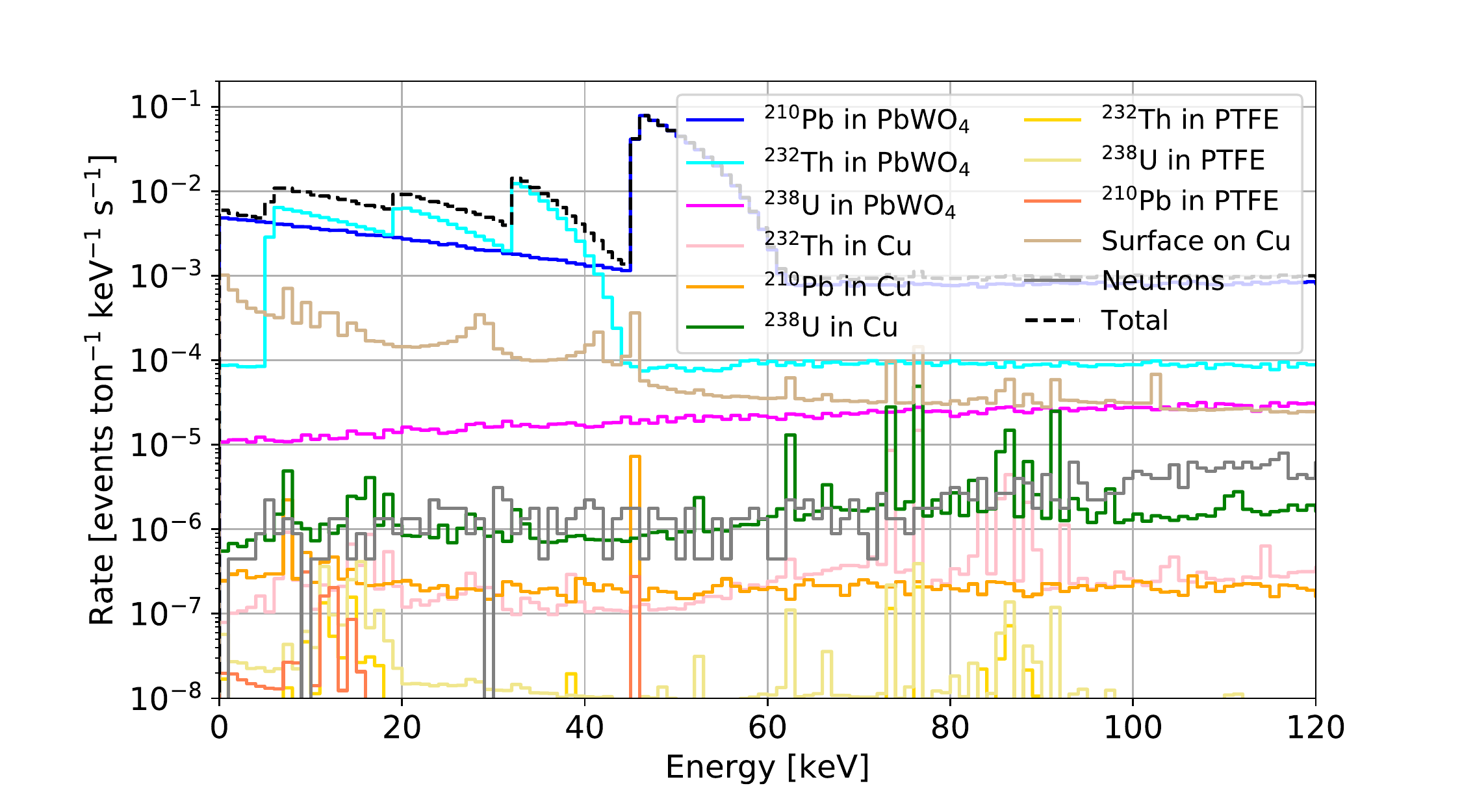}}%
\hspace{0pt}%
\subfigure[d][Accretion ${[0.1,1]}$~s - $M=4$]{%
\label{fig:CC-d}%
\includegraphics[width=0.48\textwidth]{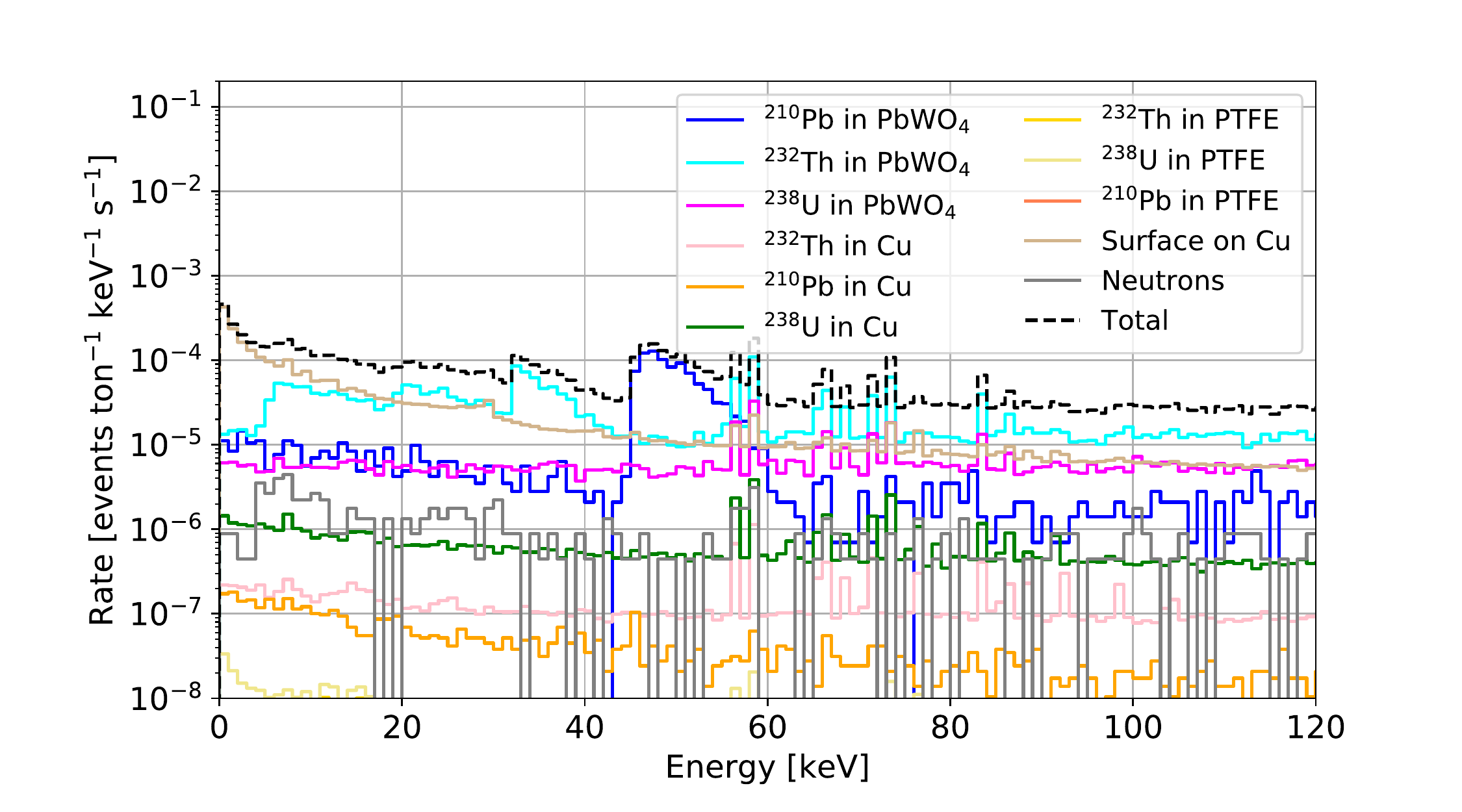}}\\

\subfigure[e][Long accretion ${[0.1,2]}$~s - $M=1$]{%
\label{fig:CC-e}%
\includegraphics[width=0.48\textwidth]{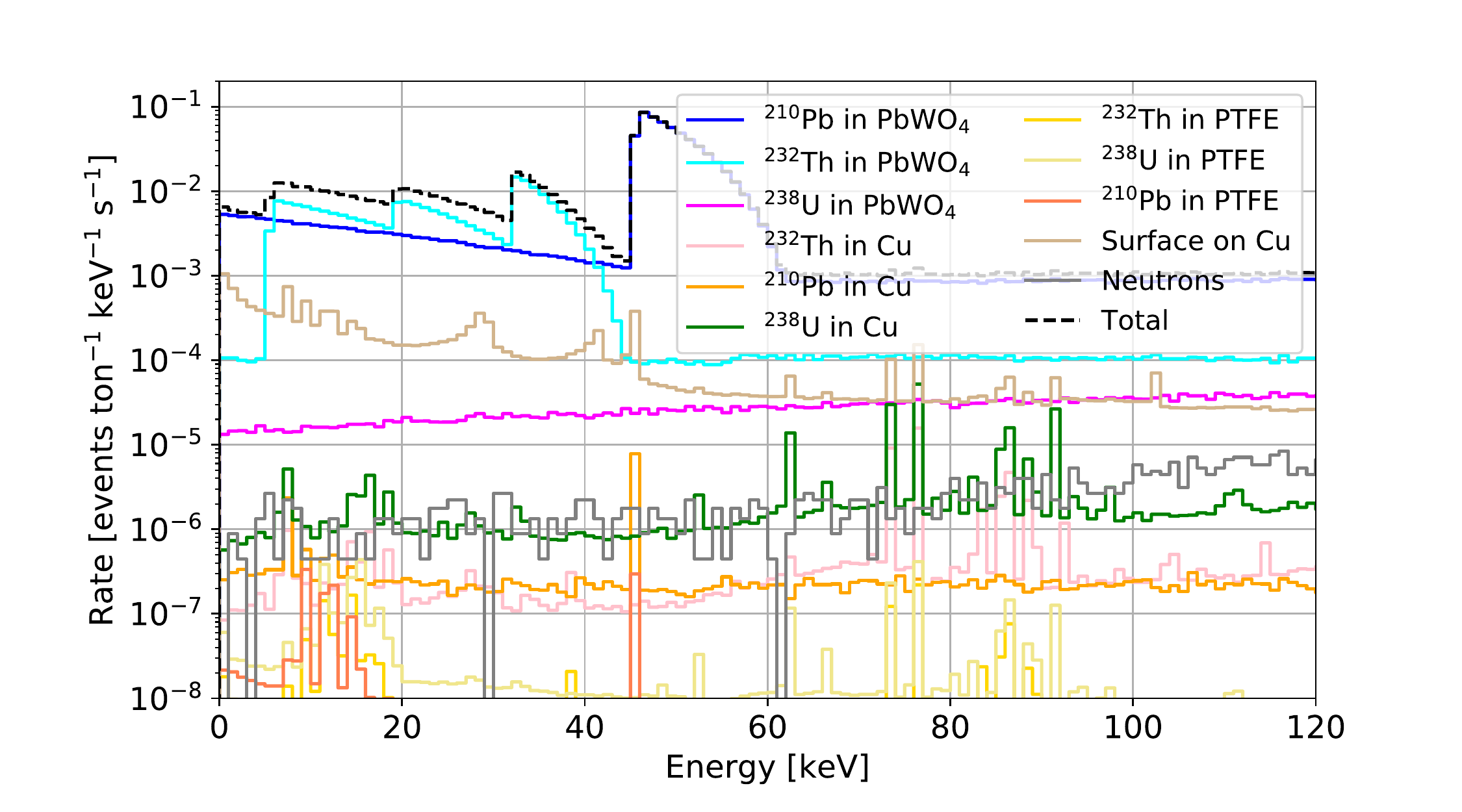}}
\hspace{0pt}
\subfigure[f][Long accretion ${[0.1,2]}$~s - $M=4$]{%
\label{fig:CC-f}%
\includegraphics[width=0.48\textwidth]{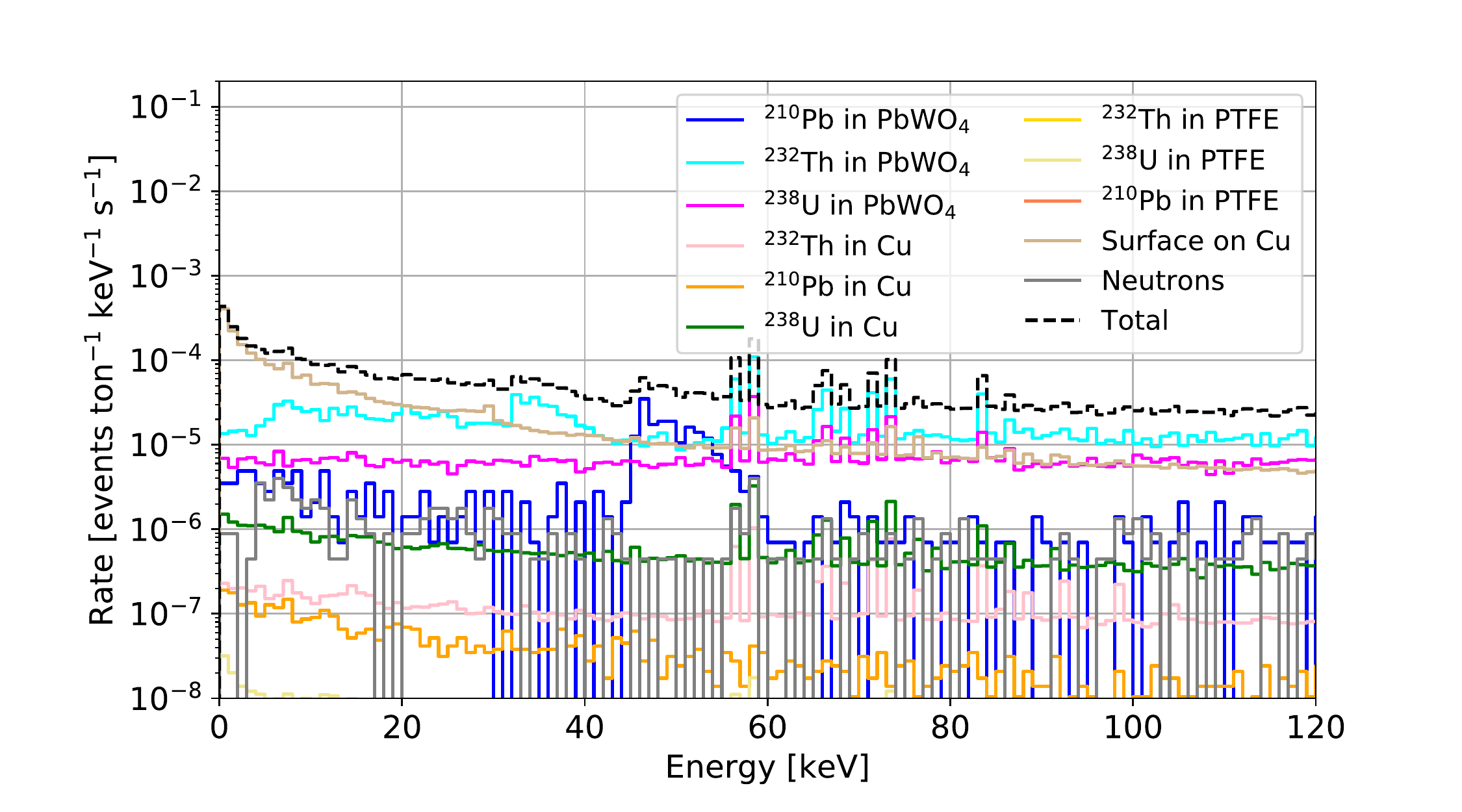}}\\
\subfigure[g][Cooling ${[1,10]}$~s - $M=1$]{%
\label{fig:CC-g}%
\includegraphics[width=0.48\textwidth]{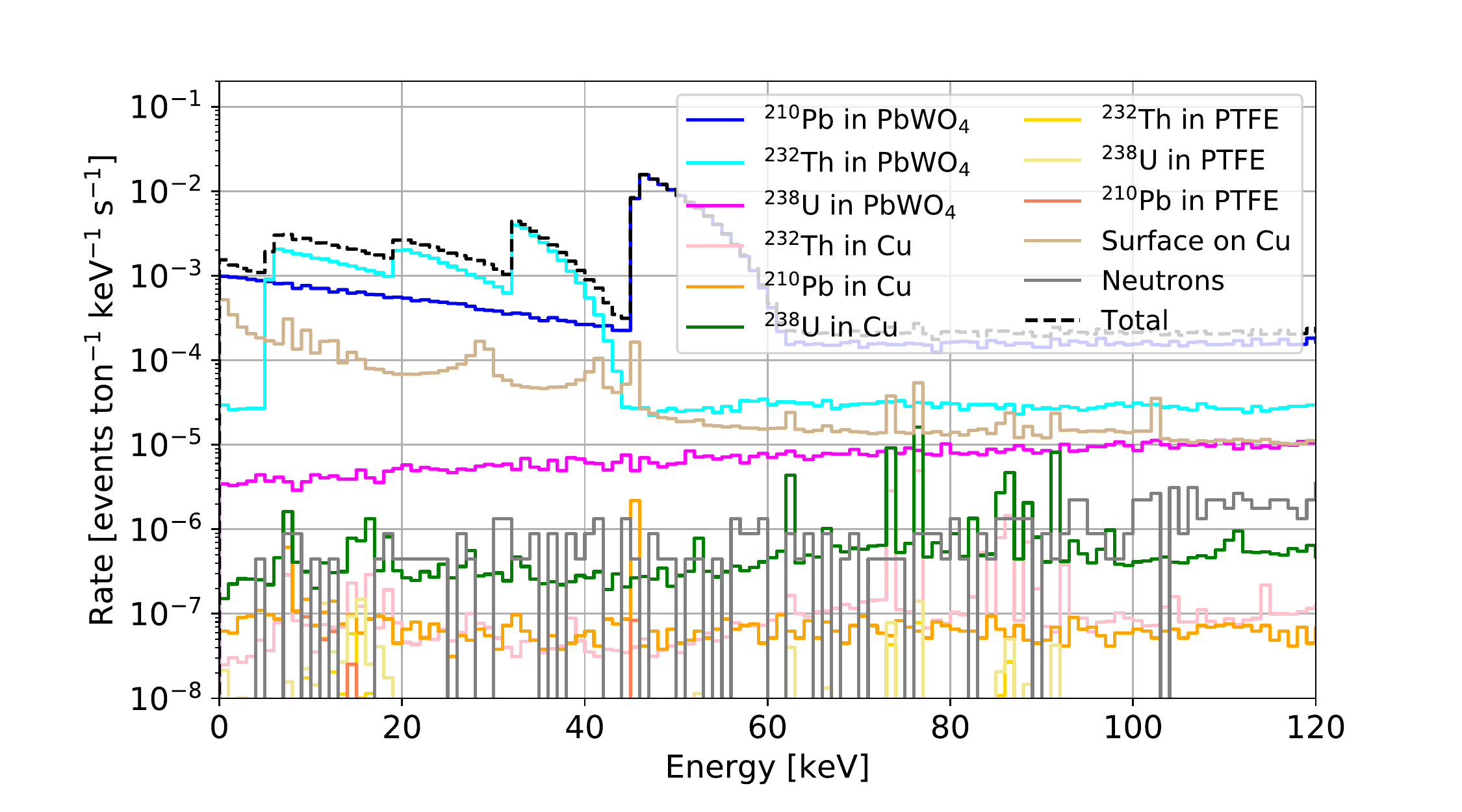}}
\hspace{0pt}
\subfigure[h][Cooling ${[1,10]}$~s - $M=4$]{%
\label{fig:CC-h}%
\includegraphics[width=0.48\textwidth]{bkg_model_M4_09s.pdf}}\\
\caption[description]{Total (nuclear and electron recoils) energy spectra produced by the background sources described in Tab.~\ref{tab:material}. Fig.~\subref{fig:CC-a}\subref{fig:CC-b},\subref{fig:CC-c}\subref{fig:CC-d},\subref{fig:CC-e}\subref{fig:CC-f},\subref{fig:CC-g}\subref{fig:CC-h} refer to the expected background rates for the neutronization ($[0.001,0.1]$~s), accretion ($[0.1,1]$~s), long accretion ($[0.1,2]$~s) and cooling ($[1,10]$~s) neutrino emission phases, respectively. In Fig.~\subref{fig:CC-a}\subref{fig:CC-c}\subref{fig:CC-e}\subref{fig:CC-g} an anti-coincidence data selection cut is applied ($M=1$), while Fig.~\subref{fig:CC-b}\subref{fig:CC-d}\subref{fig:CC-f}\subref{fig:CC-g} are coincidence spectra where 4 different detectors trigger in the same time window ($M=4$).}%
\label{fig:CC}%
\end{figure}

For the \RESNOVA detector response, we considered an energy-independent resolution $\sigma$ of 200~eV, which corresponds conservatively to an energy threshold of 1~keV, and a detector time resolution of 100~$\mu$s.

The results of the simulations are shown in Fig.~\ref{fig:CC}, where the detector energy spectra for the background sources described in Tab.~\ref{tab:material} are analyzed. These represent the expected total detector background from nuclear and electron recoils evaluated over the different neutrino emission phases. In Fig.~\ref{fig:CC}, the energy spectra for events with $M=1$, anti-coincidence spectra, and $M=4$, where any 4 of the 500 detectors are triggered in the pre-defined time window are shown. We are expecting to observe temporal correlation of background signatures (e.g. successive radioactive emissions of a decay chain), for this reason we are not expecting the background to directly scale with the detector exposure. For this reason, in the simulations we are considering the different neutrino emission phases as concatenated.

One of the most critical background source is $^{210}$Pb coming from both the crystals and the Cu structure. This can be present in different detector components as a nuclide of the $^{238}$U decay chain (bulk contamination), but also as additional independent contamination of the set-up caused for example by $^{222}$Rn implantation~\cite{Clemenza:2011zz} (surface contamination). This isotope undergoes $\beta^-$-decay with a Q-value of 63~keV, hence the electrons are in the same energy range as the expected signal. Another harmful background source is $^{228}$Ra, produced by the $^{232}$Th decay chain, which features different low energy $\beta^-$-particles (e.g. 6.6~keV, 20.2~keV and 33.1~keV) in its nuclear decay scheme~\cite{Artna-Cohen:1997uwr}.



Environmental neutrons can also interact with the detector via elastic scattering, with deposited energies inside the ROI. Given the high granularity of the detector, once a neutron enters in the set-up, it produces several interactions ($M>1$). As shown in Fig.~\ref{fig:CC}, neutrons contribute only a few percent of the background level in the ROI.

Surface contaminations on all the Cu components (Fig.~\ref{fig:CC} - \textit{surface in Cu}) give a contribution to the background in the ROI mainly at low multiplicities, because of spurious coincidences. State of the art surface purification techniques~\cite{Alessandria:2012zp} make this background source not relevant for SN neutrino investigations. 

\begin{figure}[t]
\centering
\includegraphics[width=0.7\textwidth]{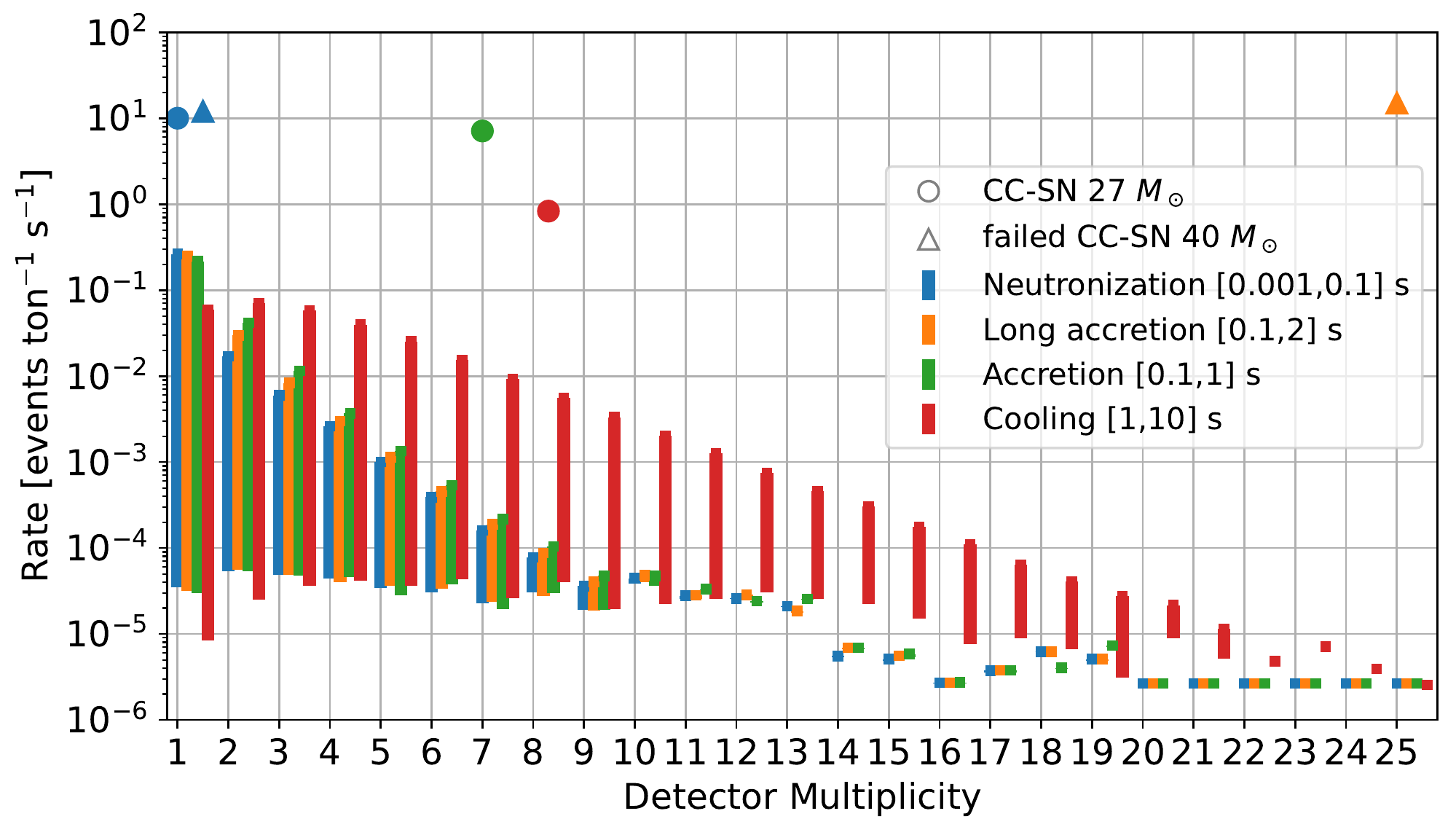}
\caption{Signal (circular and triangular markers) and background ranges (solid bars)  as a function of the number of triggered detectors (detector multiplicity). The signal values are taken from Tab.~\ref{tab:events} and normalized per time unit. The SN events are assumed to take place at 10~kpc. The background ranges are evaluated integrating the number of events over the region of interest [1-30]~keV in the energy spectra produced by the Monte Carlo simulations (see Sec.~\ref{sec:bkg} and Fig.~\ref{fig:CC}). The different colors represent the different time windows over which the signal and the background rates are computed. The bars show also the range for the expected background rate assuming a 100\% (lower value) and a null (upper value) rejection of $\beta/\gamma$ events.}
\label{fig:bkg_mult} 
\end{figure}

For the sake of comparison in Fig.~\ref{fig:bkg_mult}, we show the background counting rate in the energy range [1,30]~keV as a function of the detector multiplicity, for the different coincidence windows. As expected, the background rate increases as we largely increase the size of coincidence time window, especially for $M>1$ events. In fact, the larger the time window (e.g. cooling phase), the larger the chance of having accidental coincidences, while for narrower windows (i.e. neutronization, accretion and long accretion phases) there is no significant difference in the counting rate, having all similar lengths.

The high radiopurity level of the detector components and the nature of the simulated events (e.g. two/three-body decays, Compton scattering) make the background level decreasing as the event multiplicity increases. High multiplicity events ($M\gtrsim 5$) are ascribed to accidental coincidences.

The total background level could be further reduced for low-multiplicity events, by implementing a particle discrimination technique. In fact, PbWO$_4$ can be operated as scintillating cryogenic detector, thus enabling an identification and rejection of $\beta/\gamma$ events. In Fig.~\ref{fig:bkg_mult}, background level ranges are shown, assuming 100\% (lower end of the bar) and null (upper end of the bar) rejection of all but nuclear recoil events.

Finally, we conclude that a segmented detector has a high potential in suppressing backgrounds while searching for signals with high multiplicity, such as neutrinos from SN events. The signal-to-noise ratio changes for different signal strengths.

\section{Detector response to SN neutrinos}
\label{sec:signals}

\begin{figure}[h]
\centering
\includegraphics[width=0.7\textwidth]{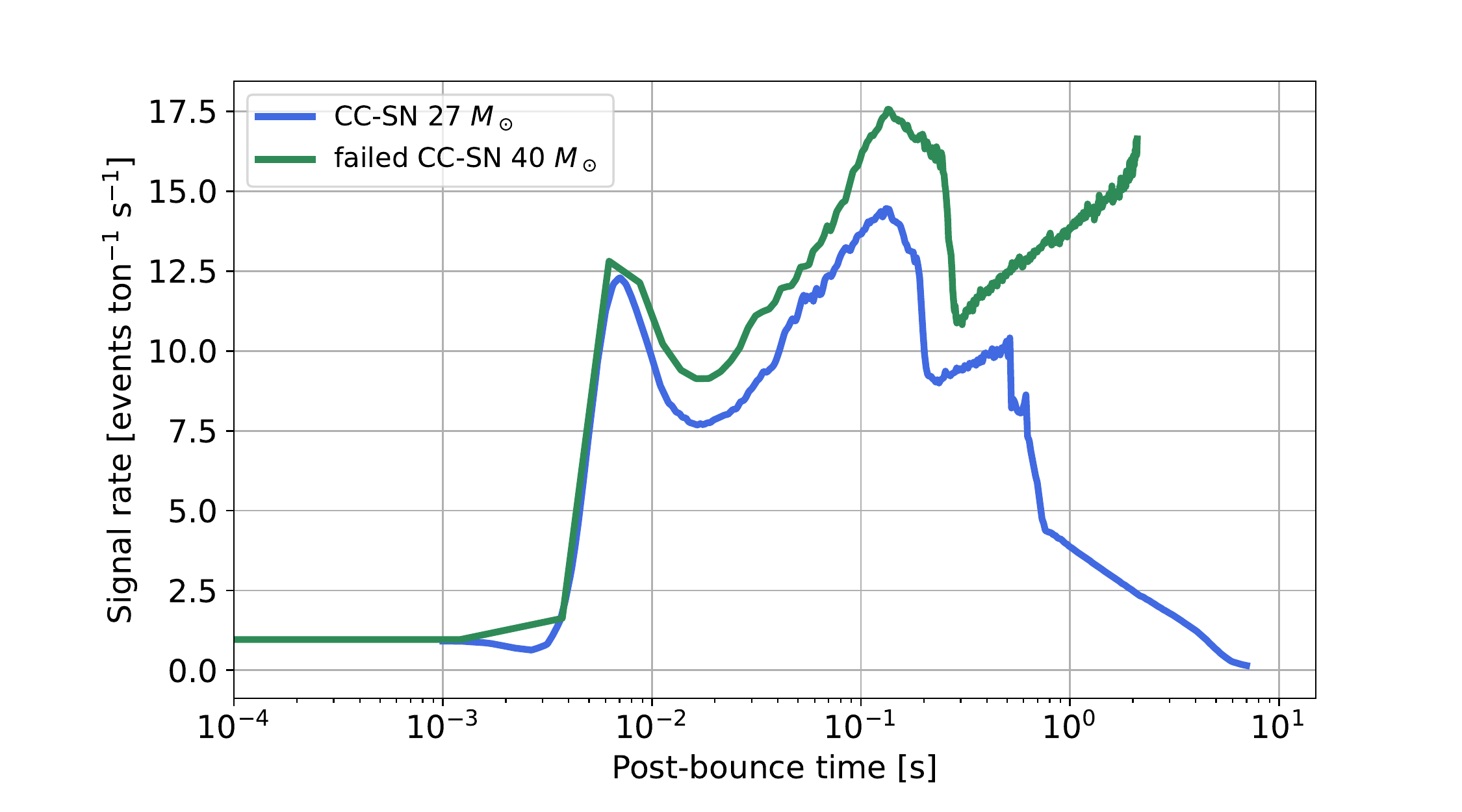}
\caption{Time response of \RESNOVA to a neutrino signal produced by a (failed) core-collapse SN with progenitor mass of $27\ M_\odot$ ($40\ M_\odot$) occurring at 10~kpc. A detector energy threshold of 1~keV is also considered.}
\label{fig:signal} 
\end{figure}

The signals produced by a core-collapse and a failed core-collapse SN shown in Fig.~\ref{fig:SN_lum} will be considered as benchmark models for our studies. The expected arrival time of neutrinos in RES-NOVA is shown in Fig.~\ref{fig:signal}.

Core-collapse neutrino signals can be easily identified from the failed core-collapse ones thanks to the change in the interaction rate during the accretion phase at times $> 0.1$~s, but also thanks to the sudden halt of the failed core-collapse signal. The difference in counting rate is due to the different progenitor masses.

The signal rate shown in Fig.~\ref{fig:signal} is computed by integrating, over the relevant incident neutrino energies, the product of the differential \CEvNS cross-section and the total neutrino fluence, and then multiplying this by the total number of target nuclei: 
\begin{equation}
\frac{dN}{dt} =\sum_{\beta} N_{\rm{PWO}} \int_{E_{\rm{min}}} f_\beta^0(E,t) \; \frac{d\sigma}{dE_R}  \;dE\ ,
\end{equation}
where $N_{\rm{PWO}}$ is the number of target nuclei, $d\sigma/dE_R$ the differential cross-section and $f^0_{\beta}(E)$ the neutrino fluence for each neutrino flavor $\beta$, and $E_{min} = \sqrt{\frac{m_N}{2}E_{thr}}$ is the minimum energy required for the incoming neutrino to induce a detectable nuclear recoil of energy $E_{thr}$ (1~keV). The neutrino fluence is obtained by integrating the total neutrino flux, computed using the so called \textit{Garching parametrization}~\cite{Keil:2002in,Tamborra:2012ac}, over the total neutrino emission time: 10~s and 2~s for the CC-SN and for the failed CC-SN, respectively:
\begin{equation}\label{eqn:flux}
f_{\beta}^0(E,t)=\frac{L_{\beta}(t)}{4 \pi d^2} \frac{\phi_\beta(E,t)}{\langle E_\beta(t) \rangle}\ ,    
\end{equation}
where $L_{\beta}(t)$ is the neutrino luminosity, $d$ the distance at which the event is occurring, $\left< E_{\beta}(t) \right>$ the average neutrino energy, and $\phi_{\beta}(E,t)$ the neutrino distribution:
\begin{equation}
\phi_\beta(E,t)=\xi_\alpha(t) \left(\frac{E}{\langle E_\beta(t)\rangle}\right)^{\alpha_\beta(t)} \exp\left(-\frac{(\alpha_\beta(t)+1) E}{\langle E_\beta(t)\rangle}\right)\ ,
\end{equation}
$\alpha_\beta(t)$ is the pinching parameter, which defines how much the distribution deviates from a perfectly thermal one, and $\xi_\beta(t)$ is obtained by $\int dE\ \phi_\beta(E,t)=1$. 

The number of expected neutrino events detected by RES-NOVA, for each phase of a SN neutrino emission window, are shown in Tab.~\ref{tab:events}.

\begin{table}[]
\centering
\begin{tabular}{r|cccc|}
\cline{2-5}
\multicolumn{1}{c|}{}                                  & \multicolumn{2}{c|}{CC SN - $27\ M_\odot$}                                                  & \multicolumn{2}{c|}{failed CC SN - $40\ M_\odot$}                      \\ \cline{2-5} 
 & {[}ev/ton{]} & \multicolumn{1}{c|}{\begin{tabular}[c]{@{}c@{}}RN\\ {[}ev{]}\end{tabular}} & {[}ev/ton{]} & \begin{tabular}[c]{@{}c@{}}RN\\ {[}ev{]}\end{tabular} \\ \hline
\multicolumn{1}{|r|}{Neutronization {[}0.001,0.1{]} s} & 1.0 & \multicolumn{1}{c|}{1.8} & 1.2 & 2.2 \\
\multicolumn{1}{|r|}{Long accretion {[}0.1,2{]} s}  & - & \multicolumn{1}{c|}{-} & 28.5 & 51.3 \\
\multicolumn{1}{|r|}{Accretion {[}0.1,1{]} s} & 7.1 & \multicolumn{1}{c|}{12.8} & - & - \\
\multicolumn{1}{|r|}{Cooling {[}1,10{]} s} & 8.3 & 
\multicolumn{1}{c|}{14.9} & - & - \\ \hline
\multicolumn{1}{|r|}{Total} & 16.4        & \multicolumn{1}{c|}{29.5} & 29.7 & 53.5 \\ \hline
\end{tabular}
\caption{Number of neutrino events detected per ton of PbWO$_4$ and by RES-NOVA (total detector mass: 1.8~ton), assuming a detector energy threshold of 1~keV. The model considered are a core-collapse SN and a failed core-collapse SN with a progenitor mass of $27\ M_\odot$ and $40\ M_\odot$, respectively. The SNe are assumed to occur at 10~kpc.}
\label{tab:events}
\end{table}

\section{\RESNOVA detection significance}
\label{sec:sens}
In order to be able to carry out a sensitivity study, where we investigate how far in space \RESNOVA can search for SNe, we need to properly estimate the background (see Sec.~\ref{sec:bkg}) and signal (see Sec.~\ref{sec:signals}) rates and their time distributions.
In addition, different statistical approaches need to be taken into account depending on relative intensity of the two. In fact, it is expected that for SN occurring at very close distances, the background is negligible compared to the large neutrino signal, however pile-up events of neutrino interactions in the detector can not be neglected. At the same time, at far distances, the neutrino signal might be as large as the background, thus another statistical approach is needed for a sensitivity estimation, possibly taking into account the possible time correlation of the neutrino events. 

In the following we show the procedure adopted for the evaluation of \RESNOVA sensitivity for two different distance ranges according to the relative values of the signal ($S$) and background ($B$) rates:
\begin{itemize}
    \item $d< 3$~kpc - the range where the detector features at least 1 pile-up neutrino event (2 neutrino events in the same crystal in less then the detector time resolution), corresponding to signal rate much greater than the background rate, $S \gg B$;
    \item $d> 3$~kpc - the range where the background rate is not negligible and the signal pile-up rate can be neglected, $S>B$; 
\end{itemize}

\subsection{SN neutrino signals at close distances ($d< 3$~kpc)}
\label{sec:near}
At close distances the number of signal events largely overwhelms the number of background events due to the large number of neutrino interactions. In this regime, the detector performance is limited by its time resolution, i.e. the maximum event rate that can be correctly resolved, and the modularity of \RESNOVA plays a key role in mitigating this issue, and in providing a precise estimation of the neutrino average energy.

Given $N_{exp}$ total expected number of neutrino events in \RESNOVA occurring in the smallest time window that a single detector can resolve, we compute the probability that two or more events pile up. First, we number the detector modules from 1 to D. Let $p_i$ be the probability for an event to occur in the $i$-th detector, the joint probability that module 1 counts $x_1$ events, module 2 counts $x_2$ events..., module D counts $x_D$ events follows the multinomial distribution:

\begin{equation}
    \begin{split}
        P(x_1,\ldots,x_D) &= \frac{N_{exp}!}{\prod_{i=1}^D x_i!}\prod_{i=1}^D p_i^{x_i} \;,\\
        \text{where} \sum_ip_i &= 1 \\
        \text{and} \sum_ix_i &= N_{exp}
    \end{split}
\end{equation}\label{eq:multinomial}
In the case of $D$ identical modules (same mass and detection threshold) all $p_i$ read $1/D$. In the case that no event occurs in pile-up all $x_i$ are 1 or 0. In addition, we disregard the particular ordering of the 1s and 0s, so we multiply by $D!$ (number of permutations of the modules) and divide by $N_{exp}!$ (number of identical 1s) and by $(D-N_{exp})!$ (number of identical 0s)\footnote{Counting of permutations with repetitions.}. Eq.~\ref{eq:multinomial} becomes:
\begin{equation}
    P(x_k \leq 1, \forall k) = \frac{N_{exp}!}{D^{N_{exp}}}\frac{D!}{N_{exp}!(D-N_{exp})!}.\label{eq:multinomialsimplified}
\end{equation}

Thanks to Eq.~\ref{eq:multinomialsimplified} we know the probability that \textit{no event} occurs in pile-up. The probability that \textit{at least} two events do pile up is the complementary of Eq.~\ref{eq:multinomialsimplified} and, making use of the shorthand notation for the \textit{falling factorial}~\footnote{ $(D)_{N_{exp}} = D(D-1)(D-2)\cdots(D-N_{exp}+1)$.}, it reads:
\begin{equation}
    P = 1 - \frac{\left( D \right) _{N_{exp}}}{ D ^{N_{exp}}}.
\end{equation}\label{eq:ppileup}
Eq.~\ref{eq:ppileup} is the probabilistic formulation of \textit{the notorious Pigeonhole principle}~\footnote{also known as Dirichlet's box principle.}~\cite{Pigeon} and represents the fraction of events that \RESNOVA cannot time-resolve.

\begin{figure}%
\centering
\subfigure[][]{%
\label{fig:pile-upcc}%
\includegraphics[width=0.48\textwidth]{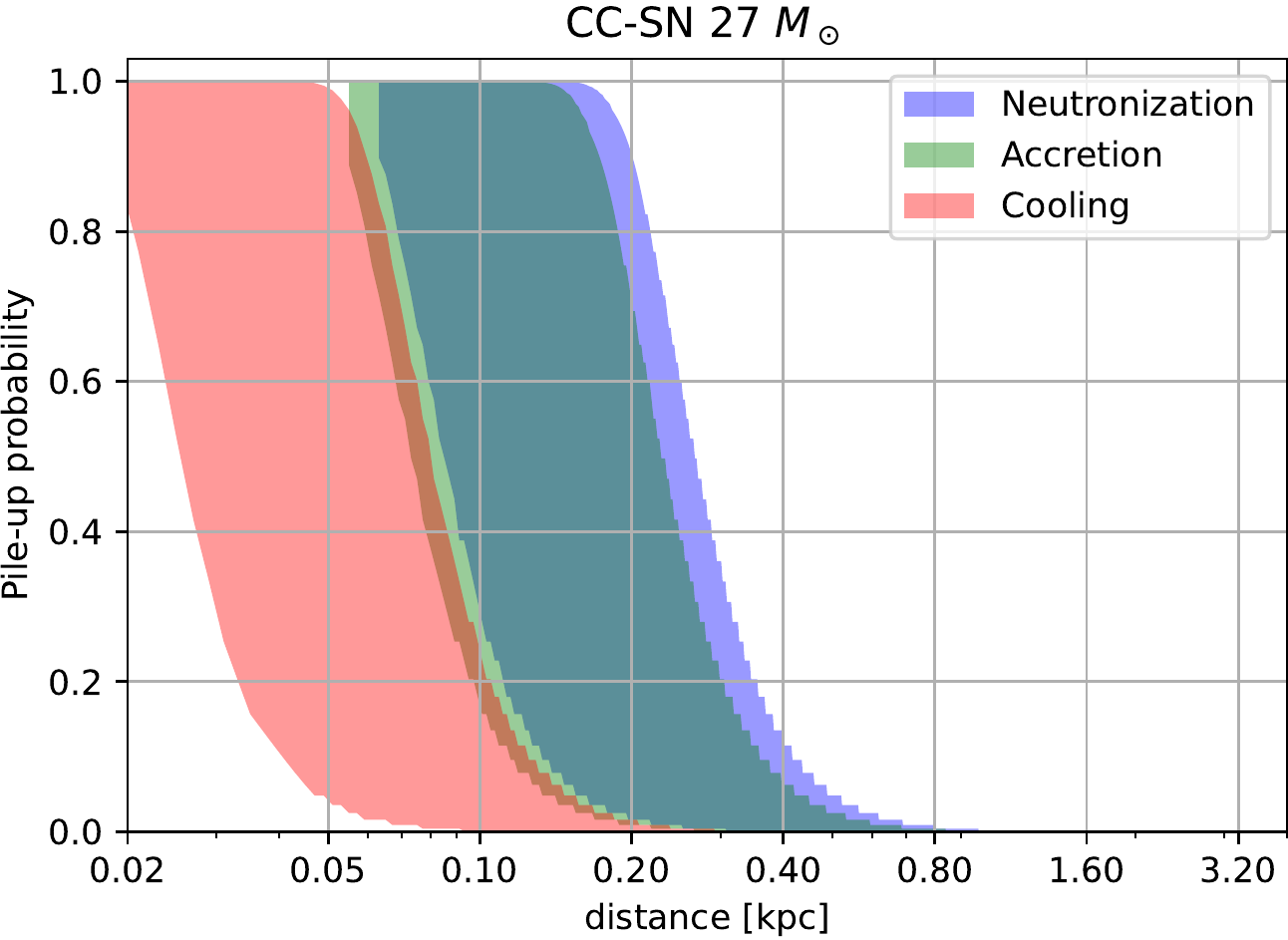}}%
\hspace{10pt}%
\subfigure[][]{%
\label{fig:pileupfailedcc}%
\includegraphics[width=0.48\textwidth]{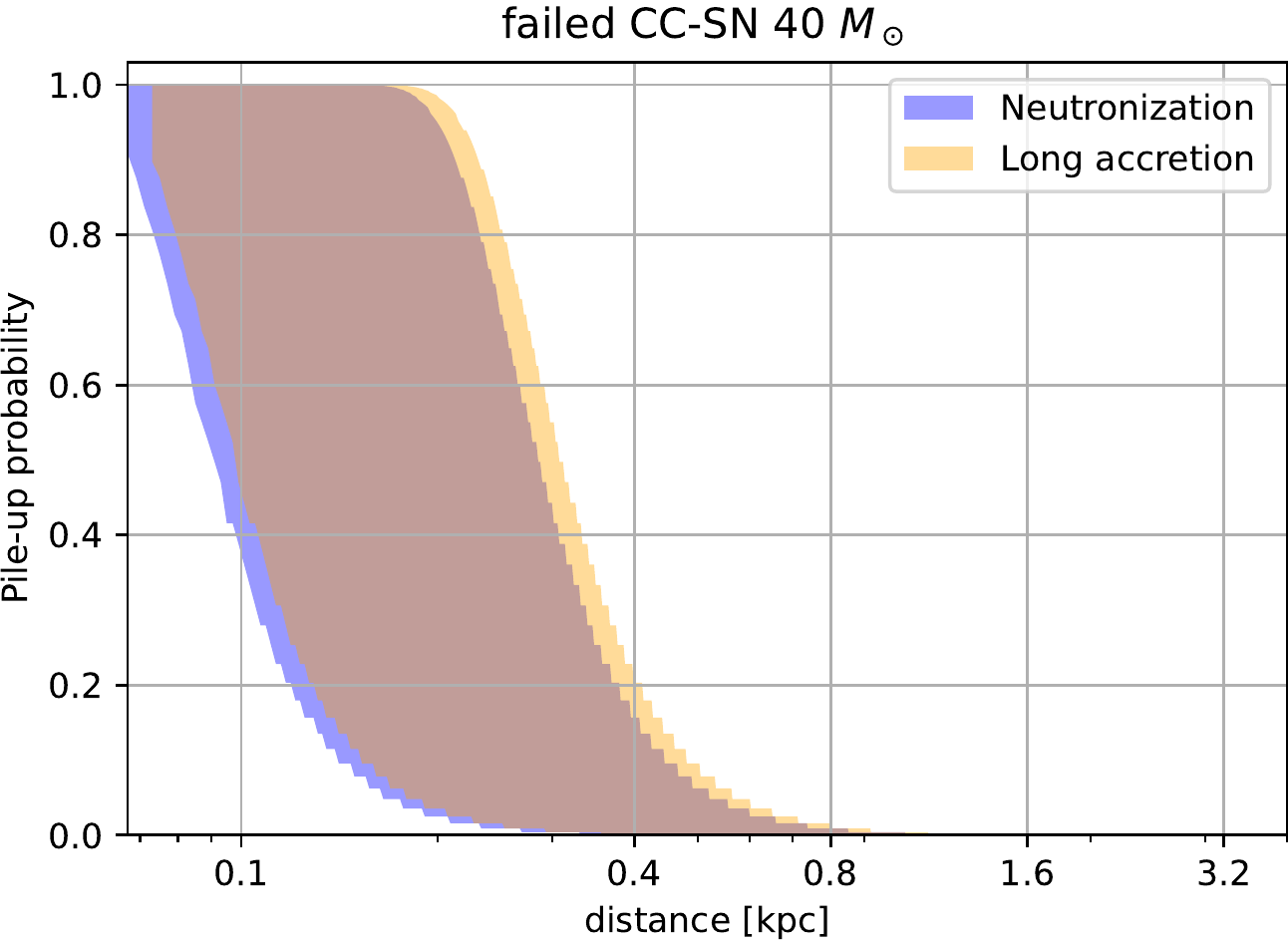}}%
\hspace{0pt}%
\caption{Range of signal pile-up probabilities in \RESNOVA as a function of the distance for different neutrino emission phases from (a) a core-collapse and (b) a failed core-collapse SN. Background sources are neglected due to their negligible contribution (see Sec.~\ref{sec:bkg}). The signal rates are taken from Tab.~\ref{tab:events}. The width of the bands represents the pile-up rejection efficiency for a detector time resolution spanning from 100~$\mu$s (best case scenario - left side of the bands) to 1~ms (worst case scenario - right side of the bands). For details on the statistical approach see Sec.~\ref{sec:near}.}\label{fig:pileup} 
\end{figure}

The high granularity of the experiment, in a conservative approach, allows to define the time resolution of the whole \RESNOVA detector with the one of a single cryogenic detector. The expected baseline value is 100~$\mu$s~\cite{Pattavina:2020cqc}, but we also considered the worst case scenario where only a time resolution of 1~ms is achieved. We compute $N_{exp}$ for the different phases of CC-SN 27 $M_\odot$ and failed CC-SN 40 $M_\odot$ as a function of the occurring distance and feed it in Eq.~\ref{eq:ppileup} to obtain $P$ for different distances. The results are shown in Fig.~\ref{fig:pileup}, where $P$ represents the probability that \textit{at least} two events can not be resolved. Assuming that all modules are equivalent, this corresponds to the fraction of events occurring in pile-up. 

\begin{table}[]
\centering
\begin{tabular}{r|c|c|ll}
\cline{2-3}
 & CC-SN $27\ M_\odot$ & \begin{tabular}[c]{@{}c@{}}failed CC-SN $40\ M_\odot$\end{tabular} &  &  \\
 & 100~$\mu$s (1~ms) & 100~$\mu$s (1~ms) &\\
 & [kpc] & [kpc] \\
 \cline{1-3}

\multicolumn{1}{|r|}{Neutronization {[}0.001,0.1{]} s} & 0.044 (0.145) & 0.047 (0.156) &  &  \\
\multicolumn{1}{|r|}{Accretion {[}0.1,1{]} s / {[}0.1,2{]} s} & 0.037 (0.125) & 0.050 (0.170) &  &  \\
\multicolumn{1}{|r|}{Cooling {[}1,10{]} s} & 0.012 (0.041) & - &  &  \\
\cline{1-3}
\cline{1-3}
\end{tabular}
\caption{Minimum distance, for each phase, for a SN event to occur that would induce a number of time-resolvable events, such that the precision in the reconstruction of the relevant parameters is no worse than in~\cite{Pattavina:2020cqc}. The values are reported for the best (worst) case scenario of 100$\mu$s (1ms) time resolution.}
\label{tab:distances}
\end{table}

In high rate conditions $S$ is so much higher than $B$ that some data loss does not spoil the sensitivity of the experiment. On the other hand, a too high rate can lead to the loss of a substantial fraction of the data. 
In order quantify the effect of the loss of data for an event at close distance, in Tab.~\ref{tab:distances} we list the distances at which the amount of neutrino events, not occurring in pile-up, is equal to the number of neutrino events expected for the same SN event at 10~kpc. This allows to specify the minimum distances at which the main parameters for each SN emission phase can be reconstructed with the same precision as claimed in~\cite{Pattavina:2020cqc}. The high detector granularity and fast time response enable to reconstruct with high precision the spectral and time features of the neutrino emission for SNe as close as Betelgeuse at 160~pc.

\subsection{SN neutrino signals in absence of pile-up ($d> 3$~kpc)}
\label{sec:far}
While considering our benchmark models, the intensity of the neutrino signal depends on the distance at which the SN is occurring and will scale as the squared distance, $d^{-2}$ (Eq.~\ref{eqn:flux}). Apparently, on the other hand the background rate should not depend on the distance at which the SN occurs. However, depending on time window over which the signal is estimated, $S(t)$, there will be a specific background rate, $B(M,t)$. The background is a function of time but also of the signal multiplicity, and so we define the multiplicity as the number of detected neutrino events $M=S(t)$. In fact, the signal and the background should be evaluated on the same detector multiplicity and the same time window.
The values for $S(t)$ at 10~kpc for each time window are shown in Tab.~\ref{tab:events}, while the $B(M,t)$ values, normalized for detector mass and window length are shown in Fig.~\ref{fig:bkg_mult}. The total number of background counts ($C_B$) during the SN neutrino emission is then computed as follow:
\begin{equation}
    C_B = m_T \cdot \sum_i B\left(S(t_i) \cdot \left(\frac{10~kpc}{d}\right)^2,t_i\right) \cdot t_i
\end{equation}
where $i$ refers to each different time window (e.g. neutronization, accretion, ...) and $m_T$ is total target mass. Background events originate from the decay chains with specific half-lives and, in the general case, $C_B$ should be considered auto-correlated in time. However, for our case of study, the background counting rate is low enough such that time-correlations do not play a noticeable role. This aspect was investigated computing the Power Spectral Density (PSD) of each decay chain, over $\approx 1$~y (3$\times 10^7$~s), and each SN model. These show no features in the signal region, therefore, we consider $C_B$ constant within each time window.

Given the low rate of SN events in the Milky Way galaxy, we assume that the background rate is measured with negligible uncertainty and is known before and after a SN neutrino burst. We determine the detection significance using the profile likelihood ratio for a simple 1-bin Poisson counting. As proven in~\cite{Cowan:2010js}, the median statistical significance can be calculated using a special, artificial data set, the \textit{Asimov data set}. Such data set is defined so that the number of signal+background events equals the number of expected signal+background events. With this method is possible to avoid large Monte Carlo simulations and evaluate the likelihoods for signal+background hypothesis and background only hypothesis on the Asimov data set (\textit{Asimov Likelihood}). The likelihood ratio is then used as test statistics to derive the detection significance. This statistical approach is the same one adopted also for sensitivity studies of other SN neutrino detection experiments~\cite{Lang:2016zhv, DarkSide20k:2020ymr,Hyper-Kamiokande:2016srs}.

\begin{figure}
\centering
\includegraphics[width=0.85\textwidth]{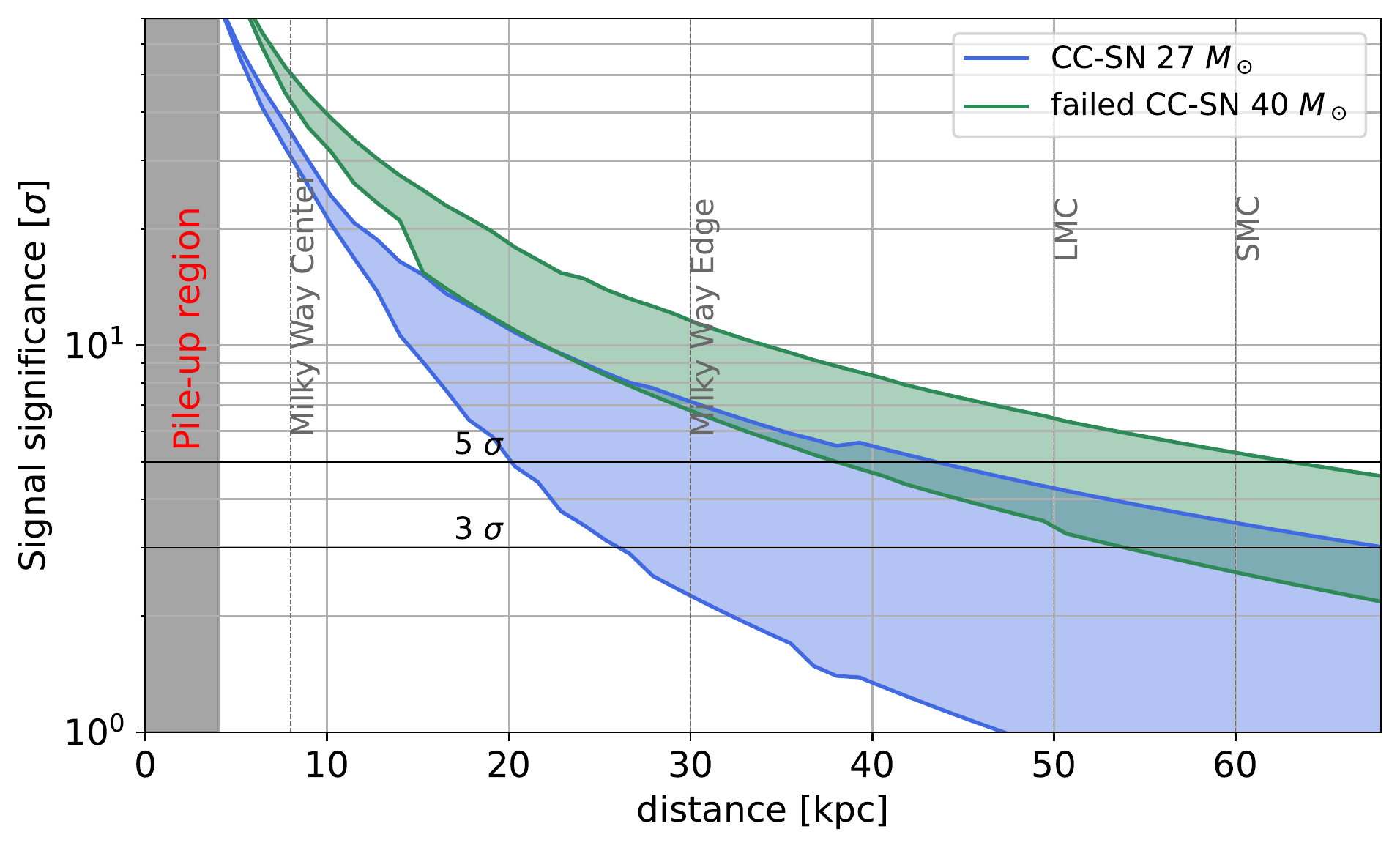}
\caption{\RESNOVA signal significance as a function of the distance for a $27\ M_\odot$ core-collapse SN and for a $40\ M_\odot$ failed core-collapse SN. The bands represent the sensitivity range achievable with (upper side of the band) and without (lower side of the band) a full-rejection of $e/\gamma$ background events. The data for significance evaluation are taken from Tab.~\ref{tab:events} (signals) and Fig.~\ref{fig:bkg_mult} (backgrounds). For details on the statistical approach, based on the Asimov data set~\cite{Cowan:2010js}, see text Sec.~\ref{sec:far}. }
\label{fig:sensitivity_medium} 
\end{figure}

In Fig.~\ref{fig:sensitivity_medium}, the signal significance is shown as a function of the distance at which the SN is occurring. The bands refer to the sensitivity range achievable with and without $e/\gamma$ background rejection. The statistical approach adopted for handling detector backgrounds in high granularity detectors is reflected on the lack of smoothness on the curves, especially for the failed core-collapse SN around 30~kpc, where the steep change in signal significance is due to a change in the background level.

The \RESNOVA detector, in the worst case scenario where no background discrimination technique is adopted, is able to probe the entire Milky Way galaxy for SN events with $>3\sigma$ signal significance. While, in the most optimistic case, with full-rejection of non-nuclear recoil events, \RESNOVA is able to reach out to the Large Magellanic Cloud with $>5\sigma$ sensitivity.
 
\section{Conclusions}
\label{sec:conc}
We have developed a detailed background model that includes contributions from the bulk and surface radioactivity of the detector, but also from environmental neutrons. These sources are expected to give the largest background contribution in the ROI and thus interfere with the search for SN events. The detector response to these sources was studied by means of conservative Monte Carlo simulations and the results have been employed to quantify the detection potential of \RESNOVA for two benchmark models, namely a core-collapse SN with progenitor mass of 27 $M_\odot$ and failed core-collapse SN with progenitor mass of 40 $M_\odot$.

We have then investigated the potential of \RESNOVA for the study of SN events in the case of nearby SN explosions (< 3~kpc), where the statistical significance is mostly limited by pile-up due to the large event-rate, and SN events at ``conventional'' distances, where the sensitivity is computed according to~\cite{Cowan:2010js}. The former case is of particular interest, since it shows the potential of the modularity of the proposed experiment given the current landscape of experimental SN neutrino searches. The unique features of \RESNOVA allow to survey SN as close as 50~pc, without substantial deterioration of the detector response due to the very high interaction rate, compared to an ideally background-, and pile-up-free equivalent experiment~\cite{Pattavina:2020cqc}.

The background budget investigated in this work lays the ground for a realistic archaeological Pb-based cryogenic detector and proves the great potential of this technology to probe for SN events on a vast range of distances, potentially from 50~pc up to the Large and Small Magellanic Clouds.

\acknowledgments
This research was partially supported by the Excellence Cluster ORIGINS which is funded by the Deutsche Forschungsgemeinschaft (DFG, German Research Foundation) under Germany’s Excellence Strategy - EXC-2094 - 390783311. We are grateful to O. Cremonesi for his valuable help in the simulation code, and to I. Tamborra for precious comments on the manuscript. We also thank R. Gaigher for his help in the detector design.



\bibliographystyle{JHEP}
\bibliography{template.bib}
\end{document}